\documentstyle[12pt,bbox]{article}
\input{epsf}
\def\beq{\begin{equation}}
\def\eeq{\end{equation}}
\def\bea{\begin{eqnarray}}
\def\eea{\end{eqnarray}}

\def\bp{{\bbox{p}} }
\def\bK{{\bbox{K}} }
\def\bq{{\bbox{q}} }
\def\br{{\bbox{r}} }

\def\boxit#1{$\phantom{a}$
\hbox{\lower0.3em\vbox{\hrule\hbox{\vrule\kern0.15em{\kern0.15em
\lower0.6em\hbox{#1}\vrule height0.3em depth 0.9em width 0pt
\kern0.15em}\kern0.15em\vrule}\hrule}}
}

\def\gappeq{\mathrel{\rlap {\raise.5ex\hbox{$>$}}
{\lower.5ex\hbox{$\sim$}}}}

\def\lappeq{\mathrel{\rlap{\raise.5ex\hbox{$<$}}
{\lower.5ex\hbox{$\sim$}}}}

\date{\today}
 
\begin{document}
 
\setcounter{page}{0}
\thispagestyle{empty}
 
\begin{flushright}
hep-ph/9811270\\
CERN-TH/98-345\\
\end{flushright}
\vspace{0.2cm}
 
\begin{center}
{\Large \bf Bose-Einstein Correlations in a Space-Time Approach to}\\
{\Large \bf $e^+ e^-$ Annihilation into Hadrons}\\
\end{center}
\bigskip
\smallskip
 
\begin{center}
 
{\large
{\boxit{\hbox{\bf K. Geiger}}\footnote{Klaus Geiger perished in the tragic
accident of Swissair flight 111 on Sept. 2, 1998. The three other
authors dedicate this common work to his memory.}}}

and

{\large
{\bf J. Ellis$^a$, U. Heinz$^{a,b}$}, and {\bf U.A. Wiedemann$^{c}$}}
\bigskip
 
$^a${\it Theoretical Physics Division, CERN,
CH-1211 Geneva 23, Switzerland\\ 
John.Ellis@cern.ch}
 
$^b${\it Institut f\"ur Theoretische Physik, Universit\"at Regensburg, 
D-93040 Regensburg, Germany\\
Ulrich.Heinz@cern.ch}

$^c${\it Department of Physics, Columbia University, New York, NY 10027, 
U.S.A.\\ awiedema@nt3.phys.columbia.edu}

\end{center}
\vspace{0.5cm}
 
\begin{center}
{\large {\bf Abstract}}
\end{center}
\smallskip
A new treatment of Bose-Einstein correlations is incorporated in
a space-time parton-shower model for $e^+e^-$ annihilation into
hadrons. Two alternative afterburners are discussed, and we use a
simple calculable model to demonstrate that they reproduce
successfully the size of the hadron emission region. One of the
afterburners is used to calculate two-pion correlations in 
$e^+e^- \rightarrow Z^0 \rightarrow hadrons$ and
$e^+e^- \rightarrow W^+W^- \rightarrow hadrons$.
Results are shown with and without resonance decays,
for correlations along and transverse to the thrust jet
axis in these two classes of events.
 
\noindent
 
\vspace{0.5cm}
 
 
\newpage

\section{Introduction}
\label{sec1} 

Perturbative parton-shower Monte Carlos~\cite{strings,clusters} combined
with models for
hadronization provide a very successful description of 
experimental data on $e^+e^- \rightarrow Z^0 \rightarrow hadrons$,
deep-inelastic lepton-nucleon scattering, etc.. In most of the
applications made so far, attention has been concentrated on
distributions and correlations in momentum space. However,
there are some key aspects of the physics where better
understanding~\cite{preconf} of the space-time development of the hadronic
system is desirable~\cite{ms37}. This is particularly true for the
treatment of dense hadronic media, such as those produced in
heavy-ion collisions, where the formation and expansion of the
system are of both experimental and theoretical interest.
A prototype for the treatment of such questions may be provided by 
the reaction $e^+e^- \rightarrow W^+W^- \rightarrow hadrons$,
where the $W^{\pm}$ do not decay independently, but in a hadronic
environment created by each other. This may engender collective
effects such as colour reconnection~\cite{GPZ,SK,GH,LL}, parton
exogamy~\cite{EG96,EG97} and Bose-Einstein 
correlations~\cite{BEtheory,LS95,LS97} that may be detected by
experiment~\cite{BEdata,LEPBE,LEP2BE}, and could be of relevance to 
the measurement of $m_W$ at LEP~2~\cite{Kunszt}.

A parton-shower Monte Carlo has recently been developed~\cite{vni} which
incorporates the information on the space-time development that 
is encoded in perturbative QCD~\cite{transport}, and combines it with a 
phenomenological spatial criterion for confinement~\cite{CEO} to provide a 
complete space-time description of hadronization. This tool has been 
applied to the analysis of $e^+e^- \rightarrow Z^0 \rightarrow
hadrons$~\cite{ms37},
$e^+e^- \rightarrow W^+W^- \rightarrow hadrons$~\cite{EG96,EG97},
deep-inelastic lepton-nucleon scattering~\cite{EGK} and relativistic heavy-ion
collisions~\cite{GHI,GM,GS}.
In the application to $e^+e^- \rightarrow W^+W^- \rightarrow hadrons$,
it has provided new insight into collective effects such as
parton `exogamy'~\cite{EG97}, namely the marriage of partons from
different $W^{\pm}$ parents to produce daughter clusters of final-state 
hadrons. In the application to relativistic heavy-ion collisions, it
has provided useful insights into such issues as the formation and 
local thermalization of the dense nuclear fireball, hadron
production~\cite{GS}, and the possible suppression of the 
$J/\psi$~\cite{GHI,GM}. However, little
attempt has so far been made to incorporate Bose-Einstein
correlations into this space-time model in a realistic way.

Bose-Einstein correlations have been analyzed in many experimental
situations, including $e^+ e^-$ annihilation~\cite{BEdata,LEPBE}
where there has also been considerable recent theoretical
progress~\cite{newBE}, and have been used extensively as a tool
to analyze the hadronic fireballs produced in relativistic heavy-ion
collisions~\cite{H96a,WTH98}. Considerable recent progress has 
been made in the development of the formalism for analyzing Bose-Einstein
correlations~\cite{H96}, and for implementing them in an afterburner for
models of hadron production~\cite{ZWSH97,Weal97,MKFW96}. It was shown 
that Bose-Einstein correlations in the two-particle momentum spectra 
allow for a detailed reconstruction~\cite{WTH98,H96} of the geometry 
and dynamical state of the reaction zone from which the final-state 
hadrons are emitted. The purpose of this paper is to describe the 
implementation of Bose-Einstein correlations in the space-time parton-shower 
Monte Carlo mentioned above (see also~\cite{WEHK98}), and to describe pilot
applications to 
the reactions $e^+e^- \rightarrow Z^0 \rightarrow hadrons$ and 
$e^+e^- \rightarrow W^+W^- \rightarrow hadrons$. This work should 
pave the way for a detailed space-time analysis of hadron production 
in these reactions using data on two-particle momentum correlations.

We introduce in Section 2 of this paper ``classical'' and ``quantum''
algorithms suitable for the calculation of Bose-Einstein correlations 
in Monte Carlo codes for hadron production. Both algorithms differ in 
how the numerical event simulation is used to define a quantum mechanical 
phase space density of emission points. We test 
these algorithms in Section 3, using a simple and analytically solvable 
model~\cite{Zajc} for the hadron source. We verify that Bose-Einstein
analysis 
tools applied to the hadronic spectra generated by the two versions 
of the afterburner reproduce correctly the input source geometry. 
Then, in Section 4 we apply the ``quantum'' afterburner to hadronic $Z^0$
and $W^+W^-$ final states generated by a parton-shower Monte Carlo. 
We calculate two-pion correlations with and without resonance decays, 
studying both the longitudinal and transverse momentum dependences of 
the correlation functions.

In a final Section, we mention possible future studies using the
approach introduced in this paper. These would include
implementation of the ``classical'' version of the Bose-Einstein
afterburner, and exploration of the influence on the ``quantum''
afterburner results of varying the assumed wave-packet size.

\section{Bose-Einstein algorithms}
\label{sec2}

In this section we explain the algorithms with which we  
later calculate two-particle correlations of identical pions from 
perturbative parton-shower Monte Carlos.

\subsection{General considerations}
\label{sec2.1}

Bose-Einstein correlations reflect the {\em phase-space} density of 
the hadronic source created in the collision. Contrary to single-particle
momentum spectra, they thus also provide access to the space-time
structure of the reaction zone. A consistent numerical simulation of 
Bose-Einstein 
effects on the two-particle and many-particle momentum distributions thus 
requires by necessity an algorithm which propagates the particles in 
phase space, rather than in momentum space only. This is what the
parton-shower cascade event generator VNI does.

The Bose-Einstein symmetrization effects result in an enhancement 
at small relative momenta $\bq$ of the 2-particle coincidence spectrum 
relative to the product of single particle spectra. The $\bq$ range of
this enhancement is inversely related to the size of the emission region
in space-time. All existing shower Monte Carlos, whether formulated
in phase space or only in momentum space, are based on a probabilistic 
description and thus do not correctly describe the many-particle
symmetrization effects of the quantum-mechanical time evolution. 
The corresponding quantum-statistical corrections must therefore be
implemented, in some approximation, by an ``afterburner'' at the
end of the classical time evolution.

In momentum-space based Monte Carlos like JETSET, one tries to implement
the clustering of identical bosons at small relative momenta by shifting 
the final state momenta according to certain prescriptions 
\cite{LS95,LS97}. These shifting prescriptions are not unique and
lead to changes in the invariant mass of the particle pair, and thus 
do not allow one to conserve simultaneously energy and momentum.
Furthermore, 
they involve a weighting function which is put in by hand but
should, in principle, reflect the (unknown) space-time structure of 
the simulated event. A recent attempt to relate the weighting 
function directly to previously unused information on the space-time 
structure of the particle-production process is described in 
\cite{AR97}. However, its connection to the position of the hadrons 
at ``freeze-out'', i.e., decoupling from the strong interactions, remains
at most indirect.

In the present paper, we study two algorithms \cite{ZWSH97,Weal97} 
to implement Bose-Einstein correlations at the end of the Monte Carlo 
simulation. These algorithms do not shift the particle momenta, nor 
do they alter the output of the event generator in any other way. 
They calculate the single-particle inclusive momentum distribution 
directly from the output momenta of the generator, and the two-particle 
coincidence spectra from the space-time coordinates and momenta of 
particle pairs from the generator output. They differ in the way
in which they associate with the event generator output a
quantum-mechanical Wigner phase space density $S(x,K)$.
Both algorithms assume that the particles propagate freely from the 
source to the detector and include only the quantum-statistical pairwise 
correlations between identical bosons. Generalizations of these algorithms
to include final-state interactions~\cite{AHR97} and multiparticle
correlation effects~\cite{W98} have been proposed but not yet
implemented numerically. 

The two-particle correlation function is constructed as the ratio of the 
two-particle coincidence spectrum $P_2(\bp_a,\bp_b)$ and the product of 
single-particle inclusive spectra, $P_1(\bp_{a,b})$,
 \begin{equation}
   C(\bq,\bK) = {\cal N}
   {P_2(\bp_a,\bp_b) \over P_1(\bp_a)\, P_1(\bp_b)}\, ,
 \label{eq1}
 \end{equation}
where $\bq = \bp_a-\bp_b$ is the relative and $\bK = (\bp_a + \bp_b)/2$ 
is the average pair momentum. With the assumption of 
independent particle emission the two-particle correlation function 
(\ref{eq1}) can then be written as \cite{S73,P84,CH94,H96}
 \begin{equation}
   C(\bq,\bK) = {\cal N}_s \left({1 + 
     {\left\vert \int d^4x\, S(x,K)\, e^{iq{\cdot}x}\right\vert^2 
      \over
      \int d^4x\, S(x,p_a)\, \int d^4y\, S(y,p_b)}
    }\right)\, ,
    \label{eq2}
 \end{equation}
where $S(x,p)$ is the single-particle Wigner phase-space density of the 
source. In this work we choose the normalization ${\cal N} = {\cal N}_s = 1$ 
in presenting our results. The implications of other choices of 
normalization are discussed in Section \ref{sec2.2}.   
The four-vectors $p_{a,b}$ in the denominator on the r.h.s. are on-shell
while the numerator contains the off-shell four-vectors $q$ and $K$ with 
$q^0=E_a-E_b$ and $K^0=(E_a+E_b)/2$. The main question 
is how to relate the event generator output to this Wigner density, 
and how to simulate the r.h.s. of Eq.~(\ref{eq2}). This will be
discussed in Section \ref{sec2.3}.

\subsection{Normalization of the correlator}
\label{sec2.2}

The normalization ${\cal N}$ in (\ref{eq1}) does not affect the 
space-time interpretation of the correlator, and the reader who is 
only interested in the latter can skip the present subsection. The 
subtle point we discuss here is that, in the context of event 
generator studies, the normalization ${\cal N}$ of the correlator 
is only fixed after requiring that the Bose-Einstein algorithm affects 
the simulated multiplicity in a particular way. We start by recalling 
the quantum field-theoretical definitions of the single- and 
two-particle spectra,
 \begin{eqnarray}
   P_1(\bp) &=& E_p \langle {\hat a}^\dagger_\bp {\hat a}_\bp \rangle\, , 
 \label{eq3} \\ 
   P_2(\bp_a,\bp_b) &=& E_a\, E_b\, \langle {\hat a}^\dagger_{\bp_a} 
       {\hat a}^\dagger_{\bp_b} {\hat a}_{\bp_b} {\hat a}_{\bp_a} \rangle \, ,
 \label{eq4}
 \end{eqnarray}
where $\langle ... \rangle$ indicates the ensemble of physical states
(events) for which the correlator is calculated. This implies the
normalizations
 \bea
 \label{eq5}
   \int P_1(\bp)\, {d^3p\over E_p} &=& \langle {\hat N} \rangle\, ,
 \\
 \label{eq6}
   \int P_2(\bp_a,\bp_b)\, {d^3p_a\over E_a}\, {d^3p_b\over E_b}
   &=& \langle {\hat N} ({\hat N} - 1) \rangle\, ,
 \eea
where ${\hat N} = \int (d^3p/E)\, {\hat a}^\dagger_\bp {\hat a}_\bp$
is the particle number operator. We now discuss the physical implications
of two different normalizations of the correlator:
 \begin{enumerate}
   \item
     One can interpret 
     the two-particle correlator as a factor~\cite{LS97}
  \begin{equation}
    d^6\sigma_{\pi\pi}^{\rm BE}/d^3\bp_1\, d^3\bp_2 
    = C(\bq,\bK)\, 
    d^6\sigma_{\pi\pi}^{\rm NO}/d^3\bp_1\, d^3\bp_2 
    \label{eq7}
  \end{equation}
     relating the measured two-particle differential cross 
     sections on the l.h.s. to the differential two-particle
     cross section resulting from the simulation. Requiring that 
     the Bose-Einstein afterburner conserves event multiplicities 
     on an event-by-event level, the corresponding momentum integrated
     total two-particle cross sections have to coincide,
     $\sigma_{\pi\pi}^{\rm BE} = \sigma_{\pi\pi}^{\rm NO}$.
     This is the appropriate starting point if total pair
     cross sections are used in the tuning of the event
     generator which then, of course, should not be changed by the 
     Bose-Einstein afterburner. 
     The normalization satisfying these requirements
     normalizes both the numerator and denominator of (\ref{eq1})
     separately to unity~\cite{GKW79},
     \begin{equation}
       {\cal N} = {{\langle \hat{N}\rangle}^2\over 
         {\langle \hat{N}(\hat{N}-1)}\rangle}\, .
       \label{eq8}
     \end{equation}
     This results in a normalization ${\cal N}_s < 1$ of the
     two-particle correlator (\ref{eq2})~\cite{MV97,W98,ZSH98}.
     \item
     A different choice of normalization often used in heavy-ion
     physics is~\cite{H96}
     \begin{equation}
       {\cal N} = {\cal N}_s = 1\, .
       \label{eq9}
     \end{equation}
     Combining (\ref{eq1}) and (\ref{eq2}), it follows that
     \begin{equation}
     \label{eq10}
     P_2(\bp_a,\bp_b) = E_a\, E_b {d^6N \over d^3p_a\, d^3p_b} 
     > P_1(\bp_a)\, P_1(\bp_b) = \left(E_a {d^3N\over d^3p_a}\right) 
     \left(E_b {d^3N\over d^3p_b}\right) \, , 
     \end{equation}
     and, because of (\ref{eq5},\ref{eq6}), also that
     \begin{equation}
     \label{eq11}
     \langle {\hat N} ({\hat N}-1) \rangle > 
     \langle {\hat N} \rangle^2\, .
     \end{equation}
     If we interpret the r.h.s. of these equations as the pair
     spectra and pair multiplicity from the event generator,
     implying that the generated multiplicity has a Poisson
     distribution, ($\langle N (N-1) \rangle_{\rm gen} = 
     \langle N\rangle^2_{\rm gen}$, then this 
     implies that the Bose-Einstein effects have increased
     the pair multiplicity. This may account for some of the 
     effects of Bose-Einstein statistics on the particle-production
     processes prior to freeze-out~\cite{B94}. 
\end{enumerate}
Depending whether we require for the Bose-Einstein afterburner the
conservation 
of event multiplicities on an event-by-event level, or aim to mimic
Bose-Einstein effects during the particle-production processes
as well, the normalization of the two-particle correlator is thus
either smaller than unity or unity itself. In the present paper, we are
only investigating the space-time interpretation of the two-particle 
correlator, and hence we can set ${\cal N} = {\cal N}_s = 1$
without any loss of generality.

\subsection{Wigner densities and event generator output}
\label{sec2.3}

We now explain how we construct a two-particle spectrum with the 
properties (\ref{eq9}-\ref{eq11}) from the event generator output. 
For simplicity, we discuss only one particle species, say $\pi^+$.
The event generator yields for each collision event $m$ a set of 
final (on-shell) $\pi^+$ momenta $p_i=(E_i,\bp_i)$ and last 
interaction points $r_i=(t_i,\bbox{r}_i)$, with $i=1,2,\dots,N_m$ 
where $N_m$ is the total number of $\pi^+$ created in event $m$:
 \beq
 \label{eq12}
     \{(r_i,p_i)\,\vert\, i=1,2,\dots,N_m\}.
 \eeq
They define a classical (positive definite) phase-space density
 \beq 
 \label{eq13}
   \rho_{\rm class}(x,p) = 
   {1\over N_{\rm evt}} \sum_{m=1}^{N_{\rm evt}}
              \rho^{(m)}_{\rm class}(x,p) =
   {1\over N_{\rm evt}} \sum_{m=1}^{N_{\rm evt}}
              \sum_{i=1}^{N_m} \delta^{(4)}(x-r_i)\,\delta^{(4)}(p-p_i)\, .
 \eeq
The distributions $\rho^{(m)}_{\rm class}(x,p)$ for individual 
events cannot be taken as Wigner densities since they fix the particle 
coordinates and momenta simultaneously, thereby violating the uncertainty
relation. This can affect the calculation of the two-particle correlator
significantly~\cite{MKFW96}. Furthermore, $\rho_{\rm class}(x,p)$ is 
always positive, whilst the Wigner density $S(x,p)$ can, at least in 
principle, become negative. Only when averaged over sufficiently large 
phase-space regions is the latter guaranteed to be positive definite. 
On the other hand, it is unlikely that
such {\em Zitterbewegung} oscillations of $S(x,p)$ or the spiky structure 
of $\rho_{\rm class}(x,p)$ affect the correlation function at small 
$\bq$ where the Bose-Einstein effects become visible. It is well known 
\cite{H96} that the width of the correlation function reflects only the 
r.m.s. width of the Wigner density $S(x,p)$ in coordinate space, and 
that finer structures in $S(x,p)$ (like spikes or quantum oscillations) 
show up in the correlator only at large $\bq$ and are very hard to 
resolve experimentally. Furthermore, since the event generator performs 
a Monte Carlo simulation of a dynamical evolution which is based on
quantum-mechanical transition amplitudes, averaging its output over
many simulated events should generate a smooth phase-space distribution 
(\ref{eq13}) which is not in conflict with the uncertainty relation.

Following these arguments, one can try to identify directly the 
classical phase-space density $\rho_{\rm class}(x,p)$ (\ref{eq13}),
averaged over sufficiently many events, with the on-shell source Wigner 
density $S(x,p)$ in (\ref{eq2}), in the following sense:
 \beq
 \label{eq14}
   \rho_{\rm class}(x,p) = 2\theta(p^0)\, \delta(p^2-m^2) S(x,p)\, .
 \eeq
This ensures the correct normalization to the average multiplicity 
$\langle N \rangle$:
 \begin{equation}
 \label{eq15}
   \int {d^3p\over E_p}\, \int d^4x\, S(x;\bp,E_p) 
   = \int d^4p \, d^4x \, \rho_{\rm class}(x,p) 
   = {1\over N_{\rm evt}} \sum_{m=1}^{N_{\rm evt}} N_m = 
   \langle N \rangle\, .
 \end{equation}
The identification (\ref{eq14}) gives rise to the ``classical'' 
version of our Bose-Einstein afterburner \cite{ZWSH97}, to be 
discussed in Sec.~\ref{sec2c1}.

Alternatively, if one wants to avoid the conceptual difficulty of 
relating an expression like (\ref{eq13}), where every term under the 
sum explicitly violates the uncertainty relation, with the source Wigner 
density, one can associate the set of phase space points (\ref{eq12}) 
with the phase-space locations of the centers of minimum-uncertainty 
wave packets~\cite{Weal97}:
  \begin{equation}
    (\bbox{r}_i, \bbox{p}_i, {t}_i) \longrightarrow
    f_i(\bbox{x},{t}_i) 
    = {1\over (\pi\sigma^2)^{3/4}}\,
      e^{-(\bbox{x}-\bbox{r}_i)^2/(2\sigma^2) 
          + i \bbox{p}_i\cdot \bbox{x} }\, .
    \label{eq16}
  \end{equation}
In this case one enforces quantum-mechanical consistency of the emission 
function $S(x,p)$ at the level of each individual simulated event.
The identification (\ref{eq16}) gives rise to the ``quantum'' version 
of our Bose-Einstein afterburner \cite{ZWSH97,Weal97}, to be discussed 
in Sec.~\ref{sec2c2}. The word ``quantum'' in this case stresses the 
quantum-mechanical consistency of the Wigner density on the {\em 
event-by-event} level (which may indeed be requiring too much), while 
the ``classical'' algorithm generates a quantum-mechanically 
consistent emission function only  on the {\em ensemble} level, and
only if $\rho_{\rm class}$ does not violate the uncertainty relation (see 
Sec.~\ref{sec3}).

Before turning to a discussion of these two algorithms we shortly comment
on the underlying assumptions. The use of 
single-particle Wigner densities $S(x,p)$ implies 
that the $N$-particle production amplitude factorizes into
one-particle production amplitudes~\cite{prattrev,H96}. 
In general, $P_2(\bp_a,\bp_b)$ is given by a sum over the two possible 
permutations of the Fourier transform of the quantum mechanical 
two-particle Wigner density $S_2(x_a,p_a;x_b,p_b)$ of the source at 
freeze-out \cite{PGG90}; here we assume $S_2(x_a,p_a;x_b,p_b) = 
S(x_a,p_a)\, S(x_b,p_b)$. This assumption (which amounts to a Wick 
decomposition of the r.h.s. of (\ref{eq4}) \cite{H96}) 
implies that the two particles in the pair are emitted independently 
from each other. It thus {\em neglects dynamical correlations} between 
the two particles in the pair, due, e.g., to energy-momentum
conservation,
as well as certain {\em quantum-statistical correlations} which may
be induced on the two-particle level by the symmetry of the multi-particle 
final-state wave function. While the neglect of dynamical correlations 
is probably well justified for heavy-ion collisions for which our algorithms 
were developed \cite{ZWSH97,Weal97}, the same is much less obvious for 
$e^+e^-$ collisions. At high energies, however, we expect such dynamical 
correlations to affect the two-particle spectrum mostly at large values 
of $\bq$, where kinematical constraints play an important role, and not 
to interfere with the Bose-Einstein correlations at small $\bq$. If this 
is true, they cancel from the ratio (\ref{eq1}) as constructed by our 
algorithm (see below). Multi-particle symmetrization effects, on the
other hand, are more of an issue in heavy-ion physics \cite{W98,CZ97}
where the rapidity densities of the produced particles are large, while
their neglect in $e^+e^-$ collisions seems unproblematic. Furthermore,
it is known \cite{ZSH98} that for certain classes of multiplicity 
distributions they do not destroy the factorization of the two-particle 
Wigner density which is assumed here.

\subsubsection{``Classical'' version of the Bose-Einstein afterburner}
\label{sec2c1}

We start from (\ref{eq13}) and (\ref{eq14}). The momenta 
returned from the event generator are on-shell, and we hence write
from now on $S(x,\bp)$ respectively $\rho_{\rm class}(x,\bp)$ for the 
on-shell distributions. The $\delta$-function structure of 
$\rho_{\rm class}$ requires one in practice to bin in the 
momentum variable (since the $x$-dependence is integrated over in
(\ref{eq2}), no binning in $x$ is necessary there). For 
this purpose we introduce the normalized ``bin functions''
with bin width $\epsilon$
 \beq
 \label{eq17}
   \delta^{(\epsilon)}_{\bp_i,\bp} = \left\{
   \begin{array}{r@{\quad:\quad}l}
     1/\epsilon^3 & 
     p_j - {\epsilon\over 2} \leq p_{i,j} \leq p_j + {\epsilon\over 2}
                \qquad (j=x,y,z) \\
     0 & {\rm otherwise}
   \end{array}
   \right.
 \eeq
or, alternatively, properly normalized Gaussians of width $\epsilon$,
  \begin{equation}
    \delta^{(\epsilon)}_{\bp_i,\bp}
    = {1\over (\pi\epsilon^2)^{3/2}}
      \exp \left( - (\bp_i - \bp)^2/\epsilon^2\right)\, .
    \label{eq18}
  \end{equation}
In the limit $\epsilon \to 0$, these Gaussian bin functions reduce to
the properly normalized $\delta$ functions 
$\delta^{(3)}(\bp_i - \bp)$.
For each event $m$ we calculate the numerator and denominator of 
(\ref{eq7}) separately. We find for the invariant two-particle spectrum 
in the numerator \cite{ZWSH97}
 \beq
 \label{eq19}
   P_2(\bp_a,\bp_b) = E_a\, E_b \sum_{i\ne j}^{N_m} \left(
   \delta^{(\epsilon)}_{\bp_i,\bp_a}\, \delta^{(\epsilon)}_{\bp_j,\bp_b}
   + \delta^{(\epsilon)}_{\bp_i,\bK}\, \delta^{(\epsilon)}_{\bp_j,\bK}
   \cos(q\cdot(r_i-r_j)) \right)
 \eeq 
and for the product of single-particle spectra in the denominator
 \beq
 \label{eq20}
   P_1(\bp_a)\, P_1(\bp_b) = E_a\, E_b \sum_{i\ne j}^{N_m}
   \sum_j^{N_m}
   \delta^{(\epsilon)}_{\bp_i,\bp_a}\, \delta^{(\epsilon)}_{\bp_j,\bp_b} \,.
 \eeq 
In (\ref{eq19}), $\bK{=}(\bp_a{+}\bp_b)/2$ and $\bq{=}\bp_a{-}\bp_b$ define 
the point in momentum space
at which the correlator is to be evaluated. Please note
that the momenta $p_{i,j}$ of the generated particles determine only
which pairs are selected and contribute to the correlator, but their 
{\em weight} in the correlator (in particular the cosine in the exchange 
term) depends only on the {\em space-time coordinates} and not on the 
momenta of the generated particles. 

The correlator (\ref{eq1}) is obtained by averaging the numerator and 
denominator separately over all events, ${1\over N_{\rm evt}}
\sum_{m=1}^{N_{\rm evt}} \dots$, and then taking the ratio. Direct 
insertion of (\ref{eq13},\ref{eq14}) 
into (\ref{eq2}) gives (\ref{eq19},\ref{eq20}) without the 
restriction $i\ne j$ on the summation indices. This is a discretization 
artefact, and the pairs with $i=j$ formed from the same particle must be 
removed by hand in this approach. 
To preserve the normalization of the correlator we 
also remove them from the denominator. 
Replacing $\cos(q\cdot(r_i-r_j))$ by $\exp(iq\cdot(r_i-r_j))$, which 
is allowed by symmetry under the exchange $i\leftrightarrow j$, the
weight function can then be factored, and we obtain
 \beq
 \label{eq21}
   C(\bq,\bK) = 1 + { \sum_{m=1}^{N_{\rm evt}} \left[
   \left\vert \sum_{i=1}^{N_m} \delta^{(\epsilon)}_{\bp_i,\bK}
              \, e^{iq\cdot r_i} \right\vert^2 
            - \sum_{i=1}^{N_m} \left(\delta^{(\epsilon)}_{\bp_i,\bK}
              \right)^2
   \right]
   \over
   \sum_{m=1}^{N_{\rm evt}} \left[
   \left(\sum_{i=1}^{N_m} \delta^{(\epsilon)}_{\bp_i,\bp_a}\right)
   \left(\sum_{j=1}^{N_m} \delta^{(\epsilon)}_{\bp_j,\bp_b}\right)
   - \sum_{i=1}^{N_m} \delta^{(\epsilon)}_{\bp_i,\bp_a}
                      \delta^{(\epsilon)}_{\bp_i,\bp_b}
   \right]}\, .
 \eeq
The subtracted terms in the numerator and denominator remove the spurious
contributions from pairs constructed of the same particles. 
The factorization of the weight function provides a dramatic
simplification.
Each of the sums in (\ref{eq21}) requires only $O(N_m)$ manipulations, 
a clear advantage for large average event multiplicities $\langle N\rangle$ 
over the evaluation of (\ref{eq19}), which involves 
$O(N_m^2)$ numerical manipulations.
Unfortunately this fails once final-state interactions are 
included, since the corresponding generalized weights no longer 
factorize \cite{AHR97}. Also, if one wants to account for
multiparticle symmetrization effects, more than $O(N_m^2)$ numerical
manipulations are typically required~\cite{W98}.

In general the result for the correlator at a fixed point $(\bq,\bK)$ 
will depend on the bin width $\epsilon$. Finite event statistics puts
a lower practical limit on $\epsilon$. 
In practice the convergence of the results must be tested numerically.
We discuss these statistical requirements in Sec.~\ref{sec3} in 
detail for a toy model.

\subsubsection{``Quantum'' version of the Bose-Einstein afterburner}
\label{sec2c2}

In the ``quantum'' version of the Bose-Einstein afterburner, the 
phase-space coordinates $(t_i,\br_i,\bp_i)$ of the generator output
are interpreted as the centers of normalized minimum-uncertainty 
Gaussian wavepackets~(\ref{eq16}) of spatial width $\sigma$.
For the one- and two-particle correlator, one 
finds instead of (\ref{eq21}) 
\cite{Weal97,ZWSH97}
 \bea
   E_p {d^3N\over d^3p} &=& {E_p \over N_{\rm evt}} 
        \sum_{m=1}^{N_{\rm evt}} \nu_m(\bp)
      = {E_p \over {N_{\rm evt}}} \sum_{m=1}^{N_{\rm evt}}
        \sum_{i=1}^{N_m} s_i(\bp)\, ,
   \label{eq22} \\
    C(\bq,\bK) &=&  1 + e^{-\sigma^2 \bq^2/2} 
     { \sum_{m=1}^{N_{\rm evt}} \left[ 
       \left\vert \sum_{i=1}^{N_m} s_i(\bK) e^{iq\cdot r_i} \right\vert^2
      - \sum_{i=1}^{N_m} s_i^2(\bK) \right]
      \over 
        \sum_{m=1}^{N_{\rm evt}} \left[ \nu_m(\bp_a)\, \nu_m(\bp_b) 
      - \sum_{i=1}^{N_m} s_i(\bp_a)\,s_i(\bp_b) \right]}\, ,
 \label{eq23} \\
   s_i(\bp) &=& \pi^{-3/2}\, \sigma^3\, 
   e^{ - \sigma^2 (\bp-\bp_i)^2}\, .
 \label{eq24}
 \eea
This result can be derived either directly from (\ref{eq16}) 
following~\cite{Weal97}, or by replacing the products of 
$\delta$ functions in (\ref{eq13})
by the Wigner densities of the corresponding wave packets, 
identifying the Wigner density $S(x,p)$
as the sum of the corresponding individual Wigner 
densities~\cite{ZWSH97,Weal97}: 
 \beq
 \label{eq25}
   \rho_{\rm class}^{(m)}(x,\bp) \mapsto S^{(m)}(x,\bp) =
   {E_p\over \pi^3} \sum_{i=1}^{N_m} \delta(x^0-t_i)\, 
   \exp\left(-{(\bbox{x}-\br_i)^2\over \sigma^2} 
             -\sigma^2 (\bp - \bp_i)^2 \right) \, .
 \eeq
In the second derivation, based on (\ref{eq25}), the spurious
contributions from identical pairs must again be removed by hand.
Now (\ref{eq25})  is correctly normalized to the number of particles 
$N_m$ in the event:
 \beq
 \label{eq26}
   \int {d^3p \over E_p} \int d^4x\, S^{(m)}(x,\bp) = N_m\, .
 \eeq
The Gaussian single particle probability $s_i(\bp)$ describes the 
contribution of the generated particle $i$ to the momentum spectrum 
at $\bp$. In the quantum algorithm, it is the counterpart of  
the bin function $\delta^{(\epsilon)}_{\bp_i,\bp}$ in 
the ``classical'' afterburner. The limit of vanishing bin width
$\epsilon \to 0$ corresponds to the limit $\sigma \to \infty$ in which
the wavefunctions (\ref{eq16}) become momentum eigenstates. The 
difference between the two algorithms is then essentially the
prefactor $\exp(-\sigma^2 \bq^2/2)$ in (\ref{eq23})
which is a genuine quantum contribution. A momentum eigenstate
is infinitely delocalized in space, and the prefactor
$\exp(-\sigma^2 \bq^2/2)$ ensures that this
infinite source size is reflected in a sharp correlator
$C(\bq,\bK) = 1 + \delta_{\bq,\bbox{0}}$. We emphasize
that while $\epsilon \to 0$ is the relevant physical limit
for the ``classical'' afterburner, $\sigma \to \infty$ is not 
the relevant limit for the ``quantum'' algorithm (\ref{eq22})-(\ref{eq24}).

It might seem natural to interpret $\sigma$ in terms of the size of 
the hadronic cluster at its formation, or of its wave function at
decoupling
from the other particles, which would suggest $\sigma$ values in the
range $\sigma \sim 1$ fm. However, such arguments are not rigorous,
and in the present paper we treat $\sigma$ as a phenomenological
parameter. One could in principle adjust this parameter as part of
a general optimization or tuning of the Monte Carlo, but such
a study extends beyond the scope of this paper.

It is important to note that, due to the smooth intrinsic momentum 
dependence of the Gaussian wavepackets (\ref{eq16}), the correlator
(\ref{eq23}) is a continuous function of both $\bq$ and $\bK$, even 
though the event generator output is discrete. On the other hand, due 
to the piecewise constant nature of the bin functions (\ref{eq17}),
the correlator (\ref{eq21}) is only a piecewise constant function of 
its arguments, which may, in practice, require binning in both $\bq$ and
$\bK$. 

The Bose-Einstein algorithms explained in section~\ref{sec2.3} 
can be applied to {\it any} model that gives a particle phase-space 
distribution, irrespective of the dynamical history of the 
particles. The aim is to reconstruct from the Bose-Einstein 
correlations information about the space-time history of the dynamical
evolution, as one attempts to reconstruct in real-life experiments
the space-time structure of collisions from the particle
distributions measured by a detector. However, with a 
particle sample from an  event generator model, detailed knowledge 
about the dynamical evolution is available. This allows one to
cross-check whether the generated dynamics reproduces the
measured Bose-Einstein correlations, i.e., this provides an
experimental test of the generated space-time 
interpretation~\cite{WEHK98}.

\section{Tests of the Bose-Einstein afterburners}
\label{sec3}

In this section we show numerical tests of our afterburners using 
a simple toy model for the source which allows for analytical
calculations of the correlation function. We thus illustrate 
the algorithms discussed in section \ref{sec2.3} before turning 
in Sec.~\ref{sec4} to realistic parton-shower calculations.

\subsection{Analytical model studies}
\label{sec3a}

We explore the above algorithms with a simple model emission function 
first proposed by Zajc \cite{Zajc}
  \begin{eqnarray}
  \label{eq27}
     \rho_{\rm class}^{\rm Zajc}(x,\bp) 
     &=& {\cal N}_s \exp \left[ -\frac{1}{2(1-s^2)} 
             \left(    \frac{\bbox{x}^2}{R_0^2} 
                 -2s\frac{\bbox{x}\cdot\bp}{R_0 P_0}
                    +  \frac{\bp^2}{P_0^2}
                 \right) \right] \delta(x^0)\, ,\qquad 0\leq s \leq 1,
    \\
    {\cal N}_s  &=& E_p\, 
    {N\over (2\pi R_0 P_0 \sqrt{1-s^2})^3} \, . 
    \label{eq28}
  \end{eqnarray}
This distribution is normalized to an event multiplicity $N$, and
is localized within a total phase-space volume 
  \begin{equation}
    V_{\rm p.s.} =  (2 R_s P_0)^3\,\qquad R_s\equiv R_0\sqrt{1-s^2}\, ,
    \label{eq29}
  \end{equation}
which vanishes for $s\to 1$. This $s$--dependence allows one to study the 
performance of our numerical algorithms for different phase-space volumes.
The parameter $s$ smoothly interpolates between completely uncorrelated 
and completely posi\-tion-momentum correlated sources: for $s \to 0$, 
the position-momentum correlation in (\ref{eq27}) vanishes, and we 
are left with two decoupled Gaussians in position and momentum space. 
In the opposite limit the position-momentum correlation is perfect, 
  \begin{equation}
  \label{eq30}
    \lim_{s\to 1} \rho_{\rm class}^{\rm Zajc}(x,\bp) \sim \delta^{(3)}
    \left(\frac{\bbox{x}}{R_0} - \frac{\bp}{P_0}\right)\,\delta(x^0)\, ,
  \end{equation}
and the phase-space localization described by the model violates the 
Heisenberg uncertainty relation. 

How are these properties reflected in the one-particle spectra and 
two-particle correlation functions? It turns out that in the Zajc model 
the two-particle correlator is independent of the pair momentum $\bK$, 
irrespective of $s$. Due to the spherical symmetry of the source and
its instantaneous time structure, the correlator is thus characterized 
by a single, $\bK$-independent Hanbury-Brown-Twiss (HBT) radius parameter.

In the ``classical'' interpretation $S(x,\bp) = \rho_{\rm class}^{\rm 
Zajc}(x,\bp)$, and the one-particle spectrum and two-particle correlator
read
 \bea
  \label{eq31}
     E_p{dN\over d^3p} &=& E_p {N\over (2\pi P_0^2)^{3/2}}\,
                        \exp\left( -{\bp^2\over 2P_0^2}\right)\, ,
 \\
 \label{eq32}
    C(\bq,\bK) &=& 1 + \exp \left( -R_{\rm class}^2\, \bq^2 \right)\, ,
 \\
 \label{eq33}
    R_{\rm class}^2 &=& R_s^2 \left( 1 - \frac{1}{(2 R_s P_0)^2}
             \right) \, .
 \eea
For sufficiently large $s$, when the phase-space volume becomes
smaller than unity,
  \begin{equation}
    s > s_{\rm crit} = \sqrt{ 1 - {1\over (2R_0P_0)^2}}\, 
    \Longleftrightarrow \, V_{\rm p.s.} < 1\, ,
    \label{eq34}
  \end{equation}
the HBT radius parameter turns negative, which leads to an 
unphysical rise of the correlation function with increasing $\bq^2$. 
Since $V_{\rm p.s.}=1$ corresponds to the volume of an elementary 
phase-space cell, the change of sign in (\ref{eq33}) is directly 
related to the violation of the uncertainty relation by the emission 
function (\ref{eq27}).

In the ``quantum'' interpretation, $\rho_{\rm class}^{\rm Zajc}$
gives the distribution of centers of Gaussian wavepackets, and the 
Wigner phase space density is obtained from $\rho_{\rm class}^{\rm Zajc}$ 
via (\ref{eq26}). The one-particle spectrum and two-particle correlator then
read
 \begin{eqnarray}
  \label{eq35}
     E_p{dN\over d^3p} &=& E_p {N\over (2\pi P^2)^{3/2}}\, 
         \exp\left( -{ \bp^2\over 2P^2}\right)\, ,
  \\
 \label{eq36}
    C(\bq,\bK) &=& 1 + \exp \left\{ -\bq^2 R_{\rm qm}^2
                         \right\}\, ,
 \\
 \label{eq37}
    R_{\rm qm}^2 &=&  R^2 \left( 1 - \frac{1}{(2 R P)^2} \right)\,,
 \\
 \label{eq37a}
    R^2 &=& R_s^2 + {\sigma^2\over 2}, \quad
    P^2 = P_0^2 + {1\over 2\sigma^2}\,.
  \end{eqnarray}
In this case, $R$ and $P$ satisfy $2RP\geq 1$ independent of the value of
$\sigma$, and
the radius parameter $R_{\rm qm}^2$ is now always positive. Even if the 
classical distribution $\rho_{\rm class}(x,\bp)$ violates the uncertainty
relation, its folding with minimum-uncertainty wave packets leads to a 
quantum-mechanically allowed emission function $S(x,\bp)$,
and to a correlator with a realistic fall-off with $\bq^2$.
The limiting cases are also as expected: For $R_0 \to \infty$
the source extends to spatial infinity and the correlator 
collapses to a Kronecker $\delta$ function at $\vert\bq\vert{=}0$. 
For $P_0 \to \infty$, the source is momentum-independent, and the 
HBT radius measures a combination of the geometric extension of 
$\rho_{\rm class}$ and the spatial wave-packet width $\sigma$:
$R_{\rm qm}^2 = R_s^2 + \sigma^2/2$. For $s=0$,
one recovers the expressions given in \cite{Weal97}. The folding 
with wavepackets modifies the geometric size of the source by adding 
in quadrature the intrinsic width of the wavepacket,
$R_{\rm intr}^2{=}\sigma^2/2$ and the size of the classical 
distribution $\rho_{\rm class}$. The extra term is exactly 
reflected by the prefactor $e^{-\sigma^2 \bq^2/2}$ by which
(\ref{eq22}) differs in structure from the classical result 
(\ref{eq22}). 

However, the spread of the one-particle momentum spectrum (\ref{eq35}) 
receives an additional contribution $1/2\sigma^2$. Choosing $\sigma$ 
too small increases this term beyond phenomenologically reasonable 
values, whilst choosing it too large widens the corresponding HBT radius 
parameters significantly. This restricts the range of phenomenologically
acceptable $\sigma$ values for the ``quantum'' version of the
Bose-Einstein afterburner. 

\subsection{Event generator studies}
\label{sec3b}

\begin{figure}[ht]\epsfxsize=13.5cm 
\centerline{\epsfbox{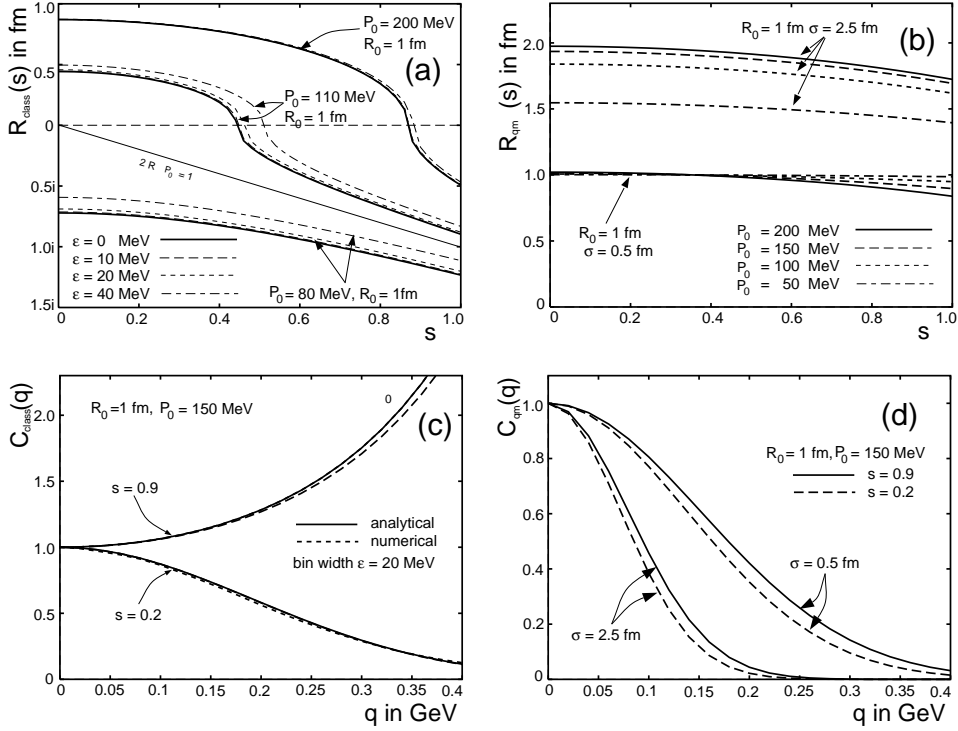}}
\caption{
Generic properties of the one-dimensional Zajc model. {\bf a)}: The 
HBT radius (\ref{eq33}) from the ``classical'' algorithm as a function of
the strength $s$ of the position-momentum correlations in the source.
Different curves correspond to different combinations of the model
parameters $R_0$ and $P_0$ and to different bin sizes $\epsilon$.
{\bf b)}: $s$--dependence of the HBT radius parameter (\ref{eq37})
for the ``quantum'' algorithm. Different curves are for different
combinations of model parameters $R_0$ and $P_0$ and different
wave packet widths $\sigma$. 
{\bf c)} and {\bf d)}: The two-particle correlator in the ``classical''
and ``quantum'' versions of the afterburner, for different sets
of model parameters. The numerical results were obtained by analyzing 
$50$ events of multiplicity $1000$, i.e. $N_{\rm pairs} = 2.5*10^7$.
They show small deviations for the ``classical'' version, but coincide
within the line widths for the ``quantum'' version.
}\label{fig1}
\end{figure}

To test the afterburner, we have mimicked the role of 
an event generator by creating a Monte-Carlo phase-space distribution 
of $N$ phase-space points $\{ (\br_i,\bp_i,t_i)\vert i=1,\dots,N\}$
according to the distribution $\rho^{\rm Zajc}_{\rm class}$
in (\ref{eq27}). This Gaussian model distribution allows one to
compare the numerical results of the Bose-Einstein algorithms
to the analytical expressions obtained above, thus testing
statistical requirements, the accuracy of the numerical 
prescriptions, and the role of the bin width in the ``classical''
algorithm. Its generic properties in both the ``classical''
and ``quantum'' versions can be read off from Fig.~\ref{fig1}.

The HBT radius parameter (\ref{eq33}) of the ``classical'' prescription,
depicted in Fig.~\ref{fig1}a, strongly depends on the position-momentum 
correlation in the source. Below $s_{\rm crit}$, it is positive,
which corresponds to a quantum-mechanically allowed Wigner function.
Above $s_{\rm crit}$, $R_{\rm class}^2$ turns negative, i.e., 
$R_{\rm class}$ 
is imaginary. The value of $s_{\rm crit}$ depends on the total
source size $2R_0P_0$ in phase space. In Fig.~\ref{fig1}a we exploited
this by varying $P_0$ between $80$ and $200$ MeV, keeping $R_0 = 1$ fm 
fixed. In the plot one sees again that the HBT radius parameter 
takes unphysical imaginary values as soon as the phase-space volume 
$(2R_sP_0)^3$ becomes smaller than 1.

As explained in Sec.~\ref{sec2c1} above, the ``classical'' Bose-Einstein 
algorithm requires a smearing of the momentum-space $\delta$ functions 
in (\ref{eq13}) by bin functions (\ref{eq17}) or (\ref{eq18}) of width 
$\epsilon$. The physical situation is recovered in the limit $\epsilon\to 0$,
but a careful investigation of this limit is numerically difficult.
However, for the Gaussian bin functions (\ref{eq18}) the HBT radius 
parameter can be obtained analytically for finite bin width $\epsilon$:
  \begin{equation}
    R_{\rm class}^2(\epsilon) = {R_s^2 \over 1 + \epsilon^2/(2P_0^2)}
    \left( 1 + {\epsilon^2 \over 2 P_0^2 (1-s^2)} 
             - {1\over (2R_sP_0)^2} \right) .
  \label{eq38}
  \end{equation}
Comparison with (\ref{eq33}) shows that the numerical results should be 
close to the physical ones if one chooses 
  \begin{equation}
    \epsilon \ll \sqrt{2}\, P_0\, .
              \label{eq39}
  \end{equation}
This provides the useful information that in practice the scale for 
$\epsilon$ is set by the width $P_0$ of the generated momentum 
distribution, independent of the geometric source size $R_s$.

In Fig.~\ref{fig1}a we have also plotted the $\epsilon$ dependence of the
HBT radius parameter. Clearly, for fixed bin width the approximation 
of the true HBT radius parameter (\ref{eq33}) becomes better with 
increasing $P_0$, as suggested by (\ref{eq38}). More generally, 
the net effect of a finite bin width is always to increase the apparent 
size of the source.

In Fig.~\ref{fig1}b we show the HBT radius obtained from the ``quantum''
version of the afterburner. Now the situation is qualitatively different:
the HBT radius is always positive, since the smearing with Gaussian wave 
packets always ensures consistency with the uncertainty relation, 
and its $s$--dependence is much weaker since the wave packets smear out 
the unphysically strong position-momentum correlations in 
$\rho_{\rm class}^{\rm Zajc}$. The different curves shown in  
Fig.~\ref{fig1}b illustrate, for fixed classical source radius $R_0$, 
the dependence of the HBT radius parameter on the width $P_0$ of the
classical momentum distribution and on the wave packet width $\sigma$.
Wave packet widths $\sigma > R_0$ not only change the HBT radius itself,
but also its dependence on $P_0$ significantly.

In Figs.~\ref{fig1}c,d we present for characteristic model parameters
the corresponding two-particle correlation functions. The analytical
curves (\ref{eq32}) and (\ref{eq36}) are compared to numerical results 
from the algorithms (\ref{eq21}) and (\ref{eq23}) applied to a 
Monte-Carlo distribution of phase-space points $\{ (\br_i,\bp_i,t_i)\vert 
i=1,\dots,N\}$ obtained from the distribution $\rho^{\rm Zajc}_{\rm class}$ 
in (\ref{eq27}). The plot shown used ${N_{\rm evt}}=50$ events of
multiplicity $N_m=1000$. We emphasize that only the total number of 
pairs in the event sample, $\textstyle{1\over 2} N_{\rm evt} N_m(N_m-1)$, 
is statistically relevant. Our choice of $N_{\rm evt}$ and $N_m$ hence 
illustrates the properties of the algorithms for both high and low 
multiplicity events.

\begin{figure}[ht]\epsfxsize=13.5cm 
\centerline{\epsfbox{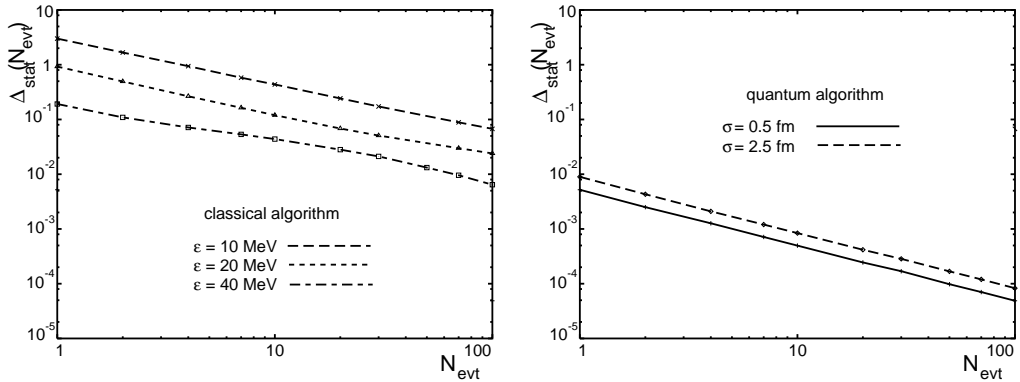}}
\caption{
The average statistical deviations $\Delta_{\rm stat}(N_{\rm evt})$ of
the ``classical'' and ``quantum'' Bose-Einstein algorithms as a function
of the number of events in the sample. Here, the event multiplicity
is $N_m = 100$, and only the total number of pairs per event sample
is statistically relevant.
}\label{fig2}
\end{figure}

For the ``quantum'' algorithm, the numerically simulated correlator
in Fig.~\ref{fig1}d coincides with the analytically calculated
one (\ref{eq36}) within the line width. Small differences between the 
analytic and numerical results are seen for the ``classical'' algorithm
in Fig.~\ref{fig1}c. In order to understand these differences in the 
performance of the two algorithms quantitatively, we have studied their 
statistical requirements in the following way: from the distribution 
$\rho_{\rm class}^{\rm Zajc}$ we generated a set of $N_{\rm fit} = 5000$ 
samples of ${N_{\rm evt}}$ events, each event containing $N_m = 100$ particles.
For each of the 5000 event samples, we calculated the two-particle 
correlator with both algorithms and determined the HBT radius 
$R_{\rm fit}$ from a Gaussian fit to this correlator. The statistical 
deviations from the analytically known exact results 
$R_{\rm class}(\epsilon)$ and $R_{\rm qm}$, respectively, were then 
determined as a function of the sample size ${N_{\rm evt}}$ via
  \begin{equation}
    \Delta_{\rm stat}({N_{\rm evt}}) = {1\over N_{\rm fit}}
    \sum_{n=1}^{N_{\rm fit}} 
    \left( R_{\rm fit}(n) - R_{\rm exact}\right)^2 \, .
    \label{40}
  \end{equation}
In general, increasing the event multiplicity $N_m$ or the number of
events ${N_{\rm evt}}$ improves the performance of the algorithms. 
Here we focus on the typical situation that the average event multiplicity
$N_m$ is fixed by the simulated physics, while the number of events
in the event sample can be increased by a longer running time of the
(numerical) experiment. The corresponding statistical performance
of both algorithms, measured in terms of $\Delta_{\rm stat}({N_{\rm evt}})$,
is plotted in Fig.~\ref{fig2}.

For the ``quantum'' algorithm, the statistical fluctuations 
decrease like $\Delta_{\rm stat}(N_{\rm evt}) \sim {1\over N_{\rm evt}}$. 
Also, their absolute value is small: for only $N_{\rm evt}=10$ events, 
the fluctuations in the fitted values $R_{\rm fit}$ are already smaller 
than 0.1 \%. This is the reason why in Fig.~\ref{fig1}d for the ``quantum'' 
algorithm the simulated values coincide so well with the analytical ones.
We also observe that $\Delta_{\rm stat}(N_{\rm evt})$ increases for larger 
values of $\sigma$. The reason is that $\Delta_{\rm stat}(N_{\rm evt}) 
\propto R_{\rm exact}^2$, which increases significantly with increasing 
$\sigma$ (see Fig~\ref{fig1}b). The normalized fluctuation measure 
$\Delta_{\rm stat}(N_{\rm evt})/R_{\rm exact}^2$ decreases
slightly with increasing $\sigma$, since the finite wave-packet width
smears out the discrete classical emission function (\ref{eq13})
and thereby reduces the statistical fluctuations in the algorithm.

In comparison, the ``classical'' algorithm shows statistical fluctuations
which are approximately two orders of magnitude larger. One sees
clearly how $\Delta_{\rm stat}(N_{\rm evt})$ increases, i.e., the
statistical requirements increase, if one goes to smaller bin widths
$\epsilon$, as needed to realize the physical limit $\epsilon \to 0$.
Also, at least for small values of $N_{\rm evt} < 100$, the
fluctuations $\Delta_{\rm stat}(N_{\rm evt})$ decrease more slowly than
$1/ N_{\rm evt}$. There are several reasons for these differences
between the ``classical'' and ``quantum'' algorithms. Numerically, we 
observe that, in the ``classical'' algorithm, the simulated correlator 
(\ref{eq21}) has even for the present Gaussian model a tendency to 
become non-Gaussian. This is seen, e.g., in the slight deviations in 
Fig.~\ref{fig1}c for $s = 0.2$. These non-Gaussian effects depend 
on $N_{\rm evt}$ and manifest themselves in the slight wiggle in 
Fig.~\ref{fig2} in the curve corresponding to $\epsilon  = 40$ MeV,
which is a relatively large bin width. Secondly, we observe that 
it is the inclusion of the Gaussian prefactor $\exp(-\sigma^2 \bq^2/2)$ 
in (\ref{eq23}) which decreases the statistical fluctuations dramatically. 
A small bin width $\epsilon$, which corresponds to a large value of 
$\sigma$, leads to large fluctuations of (\ref{eq21}), but in 
the ``quantum'' algorithm the Gaussian prefactor $\exp(-\sigma^2\bq^2/2)$ 
switches on just in the regime of ``small bin width'' and 
thereby dampens out the fluctuations.

\section{Two-Particle Bose-Einstein Correlations in a Parton-Shower Monte 
Carlo}
\label{sec4}
\begin{figure}[ht]
\begin{minipage}{80mm}
\epsfxsize=300pt
\centerline{ \epsfbox{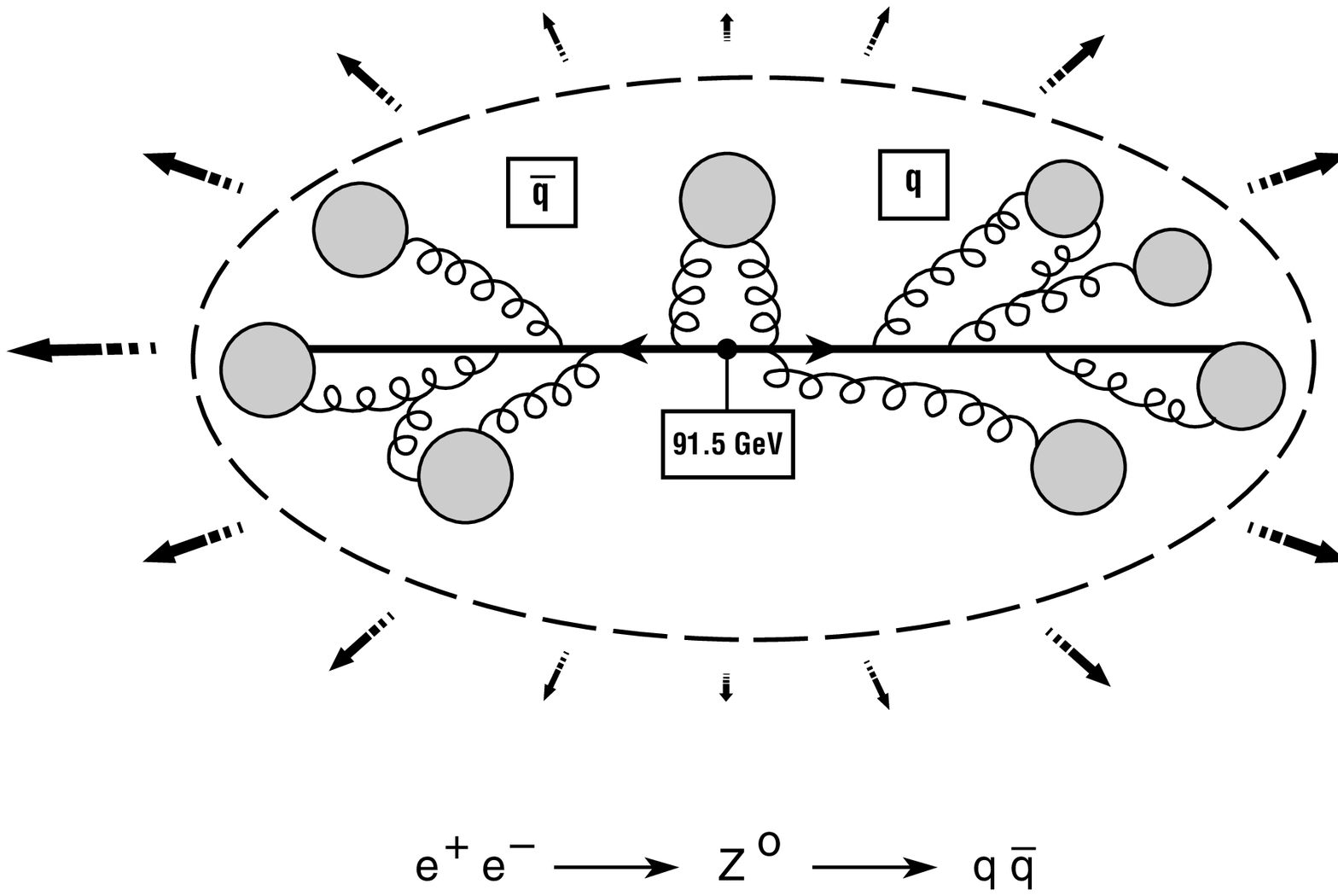} }
\end{minipage}
\begin{minipage}{80mm}
\epsfxsize=300pt
\centerline{ \epsfbox{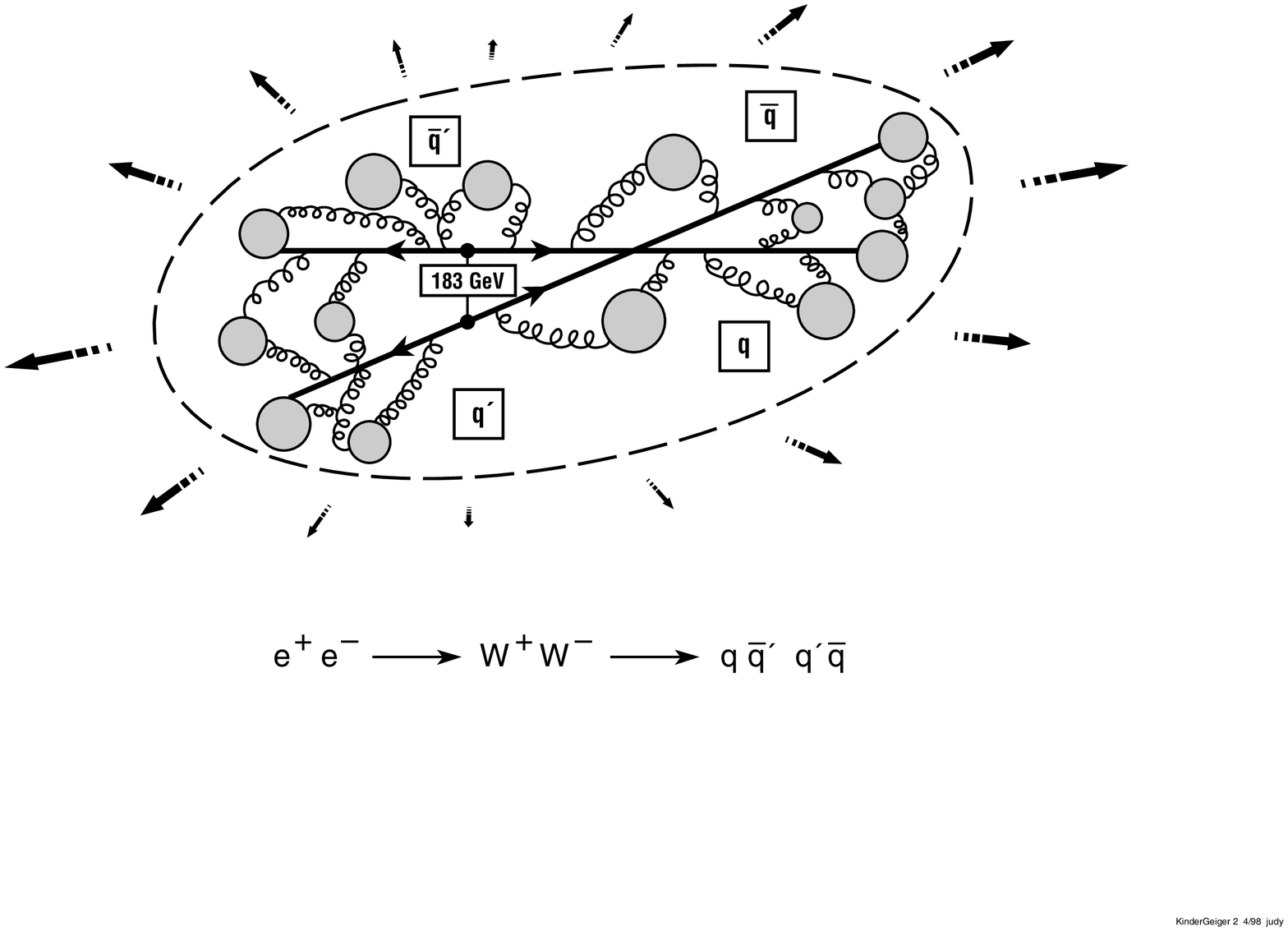} }
\end{minipage}
\caption{
Schematics of the two $e^+e^-$ event types (\ref{Z0}) and (\ref{WW}):
The final-state hadron distribution in $Z^0$ events (left) is due to 
exclusively `endogamous' hadronization of the partonic offspring from 
the $q\bar{q}$ dijet, whereas in $W^+W^-$ events (right) there is, in 
addition, the possibility of `exogamous' hadron production involving a 
mating of partons from the two different $W^+ \rightarrow q\bar{q}'$ and
$W^- \rightarrow q'\bar{q}$ dijets.
\label{fig3}
}
\end{figure}

%
Having gained some insight in the simulation of Bose-Einstein effects 
within the toy model of the previous Section, we now apply the afterburner
algorithm to the realistic case of particle emission in $e^+e^-$ 
annihilation at LEP~1~\cite{LEP1} and LEP~2~\cite{LEP2}. We focus on 
the following reaction channels, illustrated in Fig.~\ref{fig3}:
 \begin{eqnarray}
    e^+ e^- & \rightarrow &  Z^0 \; \rightarrow q\bar{q} \;\rightarrow \;
   {\rm hadrons} \quad \mbox{at $\sqrt{s} = 91.5$ GeV} \, ,
 \label{Z0}
 \\
    e^+ e^- & \rightarrow &  W^+W^- \; \rightarrow q\bar{q}' \,q'\bar{q} \;
    \rightarrow \;{\rm hadrons} \quad \mbox{at $\sqrt{s} = 183$ GeV} \, .
 \label{WW}
 \end{eqnarray}

These two processes are of interest for several reasons:
\begin{itemize}
\item
Generally, $e^+e^-$ collisions at $\sqrt(s) \ge 90$ GeV
provide the ``cleanest'' environment
of all high-energy particle collisions for studying the physics of
Bose-Einstein correlations, because there is no background
to the interesting particles emitted from the calculable parton shower,
and final-state hadrons escape unscathed from their emission point without
further interactions.
Correlation measurements can therefore be very valuable, as
they may be used to calibrate analogous analyses in 
the extreme opposite case of heavy-ion collisions, where the
emission region is more difficult to calculate accurately, and
final-state hadron scattering and cascading 
can crucially influence the shape of the particle distributions.
\item
The experimental study of $e^+ e^-\rightarrow Z^0 \rightarrow hadrons$ at
LEP~1 is based on several million events, and hence 
is impressively extensive and accurate. In particular,
high-precision measurements
of two- and three-particle correlations have been reported~\cite{BEdata}.
On the other hand, the reaction
$e^+ e^-\rightarrow W^+W^- \rightarrow hadrons$ is currently
under very active study at LEP~2, in both its experimental and
theoretical aspects: for an overview, see~\cite{LEP2}. The
interest in this reaction stems from its importance for
measuring the triple-gauge-boson couplings and $m_W$.
In particular,
it has been argued that Bose-Einstein correlations may introduce
an important source of systematic error into the analysis of $m_W$.
\end{itemize}

Theoretical studies of Bose-Einstein enhancements have mainly been 
within the context of the string models~\cite{strings}, which have been 
very successful in explaining the distributions
of identical particles seen in high-energy $e^+e^-$
collisions~\cite{BEtheory}. 
Although the string description~\cite{strings} of the hadronization
process is a very appealing phenomenological approach
and also has many other successes, it is 
not the only possible description. We employ
a rather distinct cluster hadronization model, based on a
space-time description of the perturbative development of 
parton showers, combined with a non-perturbative model for
cluster formation and hadronization~\cite{clusters}.

The crucial physics point is that, whatever model one uses for the details
of the conversion of colored partons into color-neutral hadronic
states, the Bose-Einstein correlations measured in $e^+e^-$ experiments 
are sensitive to {\it local} volumes of the order of a fermi in both 
the longitudinal and transverse directions. Therefore they provide 
important information on the intimate space-time structure of the 
hadronization mechanism. In particular, 
the sources that emit the final-state pions and other particles
must be identified with local hadronization `patches', and not with the
system as a whole, which may extend over even hundreds of fermi.
In the string picture, these local patches are the centers of
string fragments, whereas in our cluster description the patches
are elementary color-neutral clusters formed from 
the mating of nearest-neighbor partons.
The effective Bose-Einstein correlation length should correpond to the
sizes of these patches, namely the typical string extension
$\simeq 1$ fm or the mean cluster size $\simeq 0.8$ fm.
Loosely speaking, this correlation length
defines the minimum possible distance that one
may resolve from the particle distributions of the hadronic final state.
Before turning to our model-specific analysis of the Bose-Einstein
effect in $e^+e^-$ collisions, we refer the
interested reader to the comprehensive overview~\cite{haywood},
in which the status of related
experimental and theoretical research in $e^+e^-$ physics can be found.

\subsection{Modelling the space-time development of $e^+e^-$ collisions}
\label{sec41}

In order to analyze the effects of identical-particle correlations
in $e^+e^-$ collisions using the quantum version of the
Bose-Einstein afterburner, we need to concentrate on event generators that
deliver realistic, though classical, phase-space distributions of final-state
hadrons, to which we may then apply the 
afterburner simulation of quantum interference and the Bose-Einstein effect.
It is clear from the preceding Sections that such an event generator
must not only give the momentum spectra, but also the vital
space-time information on the dynamical
evolution and in particular on the final stage of hadron emission.
Unfortunately, most of the advanced event generators in particle
physics~\cite{evgens} do not encode the relevant particle emission structure
in space and time, whereas most event genarators for heavy-ion
collisions do, but cannot be applied to $e^+e^-$ physics.
One event generator that does satisfy both these requirements is
VNI~\cite{vni}, which simulates the $e^+e^-$ collision dynamics all 
the way from the hard annihilation vertex, through the perturbative QCD 
shower development to the emergence of hadronic final states.
Within the framework of relativistic quantum kinetics~\cite{ms3942},
the event generation in VNI traces in both space-time and momentum space
the parton-shower evolution from
the initial quark-antiquark pairs, followed by the clustering of the 
emitted quark and gluon offspring to
pre-hadronic cluster states that then decay into the final-state hadrons.
Referring to~\cite{EG96,vni} for details, we 
recall briefly here the essential concepts of this 
space-time model:
\begin{description}
\item[(i)]
The {\it parton-shower dynamics} is described by conventional 
perturbative QCD evolution Monte Carlo methods, with the added feature 
that we keep track of the spatial development in a series of small 
time increments. Our procedure implements perturbative QCD transport 
theory in a manner consistent with the appropriate quantum-mechanical 
uncertainty principle, incorporating parton branching due to real and 
virtual quantum corrections involving gluons or quark-antiquark pairs.
In the rest frame of the $Z^0$ (for (\ref{Z0})) or of the $W^{\pm}$ 
(for (\ref{WW})), each off-shell parton $i$ in the shower propagates for 
a time $\Delta t_i$ given in the mean by $<\Delta t_i> = \gamma_i 
\tau_i = E_i /k^2_i = x_i M /2 k^2_i$, where $k^2_i$ is the parton's 
squared-momentum virtuality, and $x_i =  E_i / M$ is its longitudinal 
energy fraction, during which it travels a distance $\Delta r_i = 
\Delta t_i \beta_i$, where $M = M_{Z^0}$ for (\ref{Z0}) and $M=M_W$ 
for (\ref{WW}). It has been shown~\cite{EG96} that such a description 
results in a typical inside-outside perturbative cascade~\cite{marchesi}.
\item[(ii)]
The {\it parton-hadron conversion} is handled using a {\em strictly 
spatial} criterion for confinement, with a simple Ansatz for the 
probability $P(r)$ that nearest-neighbor color charges coalesce to 
color-neutral clusters in accord with their color and flavor degrees 
of freedom, where $r$ is the relative distance in between them in their
center-of-mass frame. The nearest-neighbor criterion is imposed 
at each time step in the shower development, in such a way that every
parton that is further from its neighbors than a certain critical
distance $R_c = 0.8$ fm has a probability distribution smeared around 
$R_c$ for combining with its nearest-neighbour parton to form a 
pre-hadronic cluster, possibly accompanied by one or more partons to 
take correct account of the colour flow. It is important to stress that 
at no moment in this shower development do we incorporate
any prejudice regarding the genealogical origin of the partons:
an `exogamous'~\cite{EG97} pair of partons from different mother
$q {\bar q}'$ pairs have the same
probability of coalescing into a hadronic cluster as do an `endogamous'
pair of partons from the same mother, at the same spatial separation.
The resulting hadronic clusters are then allowed to decay into
stable hadrons according to the particle data tables.
\end{description}

In the present paper, due to the less demanding requirements on event 
statistics, we will use the ``quantum'' version of the Bose-Einstein 
afterburner. This requires fixing the wave packet width $\sigma$ in the
algorithm. Lacking convincing arguments for a unique physical choice
of this parameter, we try to connect it with the intrinsic size $R_c$ 
of the pre-hadronic clusters which act as pion-emitting sources. 
The size $R_c$ also defines the minimum distance of adjacent clusters 
at formation without overlapping. We thus set
 \begin{equation}
   R_c = 0.8\; \mbox{fm} = \sigma\, .
 \label{Rc}
\end{equation}
Finally, we stress that, although
the event generation of $e^+e^-$ collisions along the above lines
should provide a rather realistic simulation of the particle dynamics,
we do not claim our results to be more than qualitative at this point, 
mainly since we have not included final-state interactions
among the produced hadrons due to either Coulomb or strong interactions.
In our approximation, the production vertex 
of each final-state hadron marks the
last point of interaction, beyond which the particles stream freely 
on classical trajectories.
Since in `real-life' experiments these final-state interactions can 
become large at small relative momenta, one should 
be careful when comparing to measured data, some of which have
been corrected for final-state interactions, others not.

\subsection{Results for two-pion correlations}
\label{sec42}
\subsubsection{Multiplicity distributions}
\label{sec421}

\begin{figure}[ht]\epsfxsize=9.5cm 
\centerline{\epsfbox{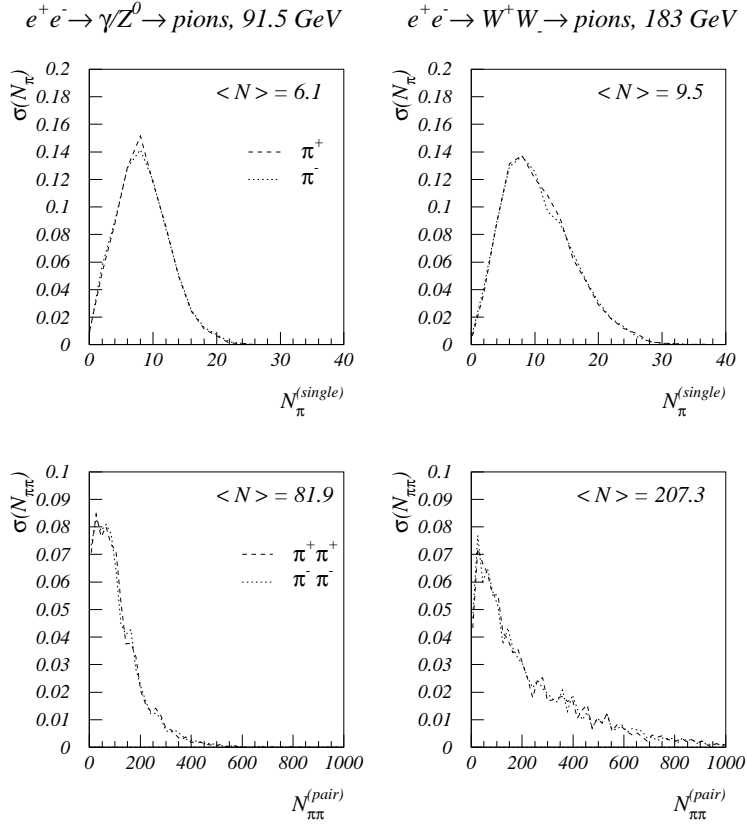}}
\caption{
Multiplicity distributions of single pions (top) and of pairs of 
identical pions (bottom), per charged pion species.
}\label{fig4}
\end{figure}

It is plausible that the structure of the hadronic final state
in $e^+ e^-  \rightarrow   W^+W^- \rightarrow q\bar{q}' \,q'\bar{q}$
may not merely be a copy of $W^{\pm}  \rightarrow q\bar{q}'$ with 
twice the final-state multiplicity. As discussed in detail 
in~\cite{EG97,EG96}, it was found within our space-time parton-shower model
that not only  the total multiplicity $N(W^+W^-)$ may be smaller than
$2\times N(W^{\pm})$, but also that the particle spectra may exhibit
characteristic differences. These
differences are due to the special geometry of $W^+W^-$ events,
in which the partonic offspring of the $W^+$ dijet overlap
in space-time with the partons emitted from the $W^-$ dijet.
The cross-talk between the quanta from the $W^{\pm}$ is especially
prominent at small rapidities and if the two dijets emerge at small
relative angles. Then, whereas in 
$e^+e^- \rightarrow Z^0$ decays all particles come from the same
mother and only `endogamous' cluster formation is possible,
as in the left part of Fig.~\ref{fig3},
$W^+W^-$ events receive a significant contribution from
the coalescence of partons from different $W^{\pm}$ mothers into 
`exogamous' clusters, as in the right part of Fig.~\ref{fig3}.

Fig.~\ref{fig4} reflects the effects of parton `exogamy' in 
$e^+ e^-  \rightarrow   W^+W^- \rightarrow q\bar{q}' \,q'\bar{q}$
as compared to $Z^0  \rightarrow q\bar{q}$, in the multiplicity
distributions of both single pions (top) and of pairs of identical 
pions (bottom). Even allowing for the slightly larger mass of the $Z^0$
compared to the $W^{\pm}$, one observes, in agreement with the above 
discussion, an effective reduction $N_{\pi^\pm}(W^+W^-) < 2 
\times N_{\pi^\pm}(Z^0)$, namely,
$\langle N_{\pi^\pm}(Z^0) \rangle= 6.1$ versus
$\langle N_{\pi^\pm}(W^+W^-)\rangle = 9.5$, per pion species.
Similarly, we find for the number of identical-pion pairs
$N_{\pi^\pm\pi^\pm} = N_{\pi^\pm} ( N_{\pi^\pm}  - 1)$ that
$N_{\pi^\pm\pi^\pm}(W^+W^-) < 4 \times N_{\pi^\pm\pi^\pm}(Z^0)$,
namely, $\langle N_{\pi^\pm\pi^\pm}(Z^0) \rangle= 81.9$ versus
$\langle N_{\pi^\pm\pi^\pm}(W^+W^-) \rangle= 207.3$.

The effect may be thought of as reflecting increased `efficiency' in the
hadronization process, due to the fact that the presence of two 
cross-talking dijets in the $W^+W^-$ decays,
with their spatially-overlapping offspring, allows
the evolving particle system to reorganize itself more favorably in
the cluster-hadronization process, and to form clusters with smaller 
invariant mass than in the $Z^0$ events. Indeed,
it was found in~\cite{EG97} that the mass spectrum of pre-hadronic
clusters from coalescing partons is in fact softer in the
$W^+W^-$ case, reflecting the fact that the availability of
more partons enables clusters to form from
configurations with lower invariant mass than
in the $Z^0$ case. 
\subsubsection{Origins of pions}
\label{sec422}

In theory, all pairs of identical pions can exhibit Bose-Einstein
correlations. Experimentally, however, the measurements of the pair
spectrum in the relative pair momentum $q$ run out of statistics because
the phase space vanishes at very low $q$.  Since small $q$ values
correspond to large spatial distances, this region of $q$ is particularly
sensitive to the decays into pions of long-living resonances, and also to
long-range Coulomb or strong final-state interactions among the particles. 
Whereas final-state interaction effects can be corrected, this is not easy
for resonance decays.  Since many of the pions in $e^+e^-$ collisions have
their origins in the decays of other particles with lifetimes
significantly greater than a few fermis, it is useful to disentangle the
various experimental sources of pions (or other particles) and to classify
their parents as follows~\cite{haywood,BEresonances,ursuli}:  
\begin{itemize} 
\item {\sl Prompt production} leading to pions that emerge directly from
the hadronization of the fragmenting system, whose parents may be visualized
as decaying strings or (in our case)  as pre-hadronic clusters.  
\item
{\sl Short-lived particles} such as $\rho$, $K^\ast$ and $\Delta$, that
are strongly-decaying particles with decay lengths shorter than a few
fermis.  
\item {\sl Long-lived resonances}, such as $\eta,\eta ', \omega,
\phi$, that are states which also decay strongly but have life-times of
many fermis.  
\item {\sl `Stable' particles}, such as $\Lambda$ and
$K_s^0$, that are particles which propagate sufficiently far that the
pions emerging can be removed by track cuts.  
\item {\sl Weakly-decaying particles}, such as charm or bottom mesons.  
\end{itemize}

\begin{table}[ht]
\begin{center}
\begin{tabular}{|c|cc|}
\hline\hline
& & \\
$\;\;\;\;\;\;\;\;$ origin $\;\;\;\;\;\;\;\;$
&
$\;\;\;\;\;$ life-time $\tau$ $\;\;\;\;\;$
&
$\;\;\;\;\;$ fraction $\;\;\;\;\;$
\\
& & \\
\hline\hline
& & \\
clusters      & $<$ 0.5 fm &  0.31  \\
$\rho, \Delta, K^\ast$   & 1.3 $-$ 4 fm &  0.41  \\
$\eta, \eta', \omega, \phi$   & $>$ 10 fm &  0.28  \\
& & \\
\hline\hline
\end{tabular}
\end{center}
\caption{
Relative contributions of different sources of pions in our 
$e^+e^-$ event simulations.
\label{T1}}
\end{table}

In Table~\ref{T1}, we list the fractions of pions coming from
these different sources, as estimated in our model simulations.
Since the `stable' particles can be considered as not having decayed,
and weakly-decaying particles contribute only a negligible fraction,
we do not include these two categories in the list.
We observe that the numbers in Table~\ref{T1} are very similar
to those reported in~\cite{haywood}.

\subsubsection{The pion correlator {\boldmath $C(q,K)$} for 
{\boldmath $K = 0$}}
\label{sec423}

Fig.~\ref{fig5} shows the correlator $C(\bq,\bK) - 1$ for different 
$\bq$-values and vanishing pair momentum $\bK$ in the c.m. frame of 
the collision, $C(q_z,q_s,q_o,\bbox{0}) - 1$. Two interesting observations 
can be made immediately:

\begin{figure}[ht]
\vspace*{1cm}
\epsfxsize=9.5cm 
\begin{minipage}{80mm}
\centerline{ \epsfbox{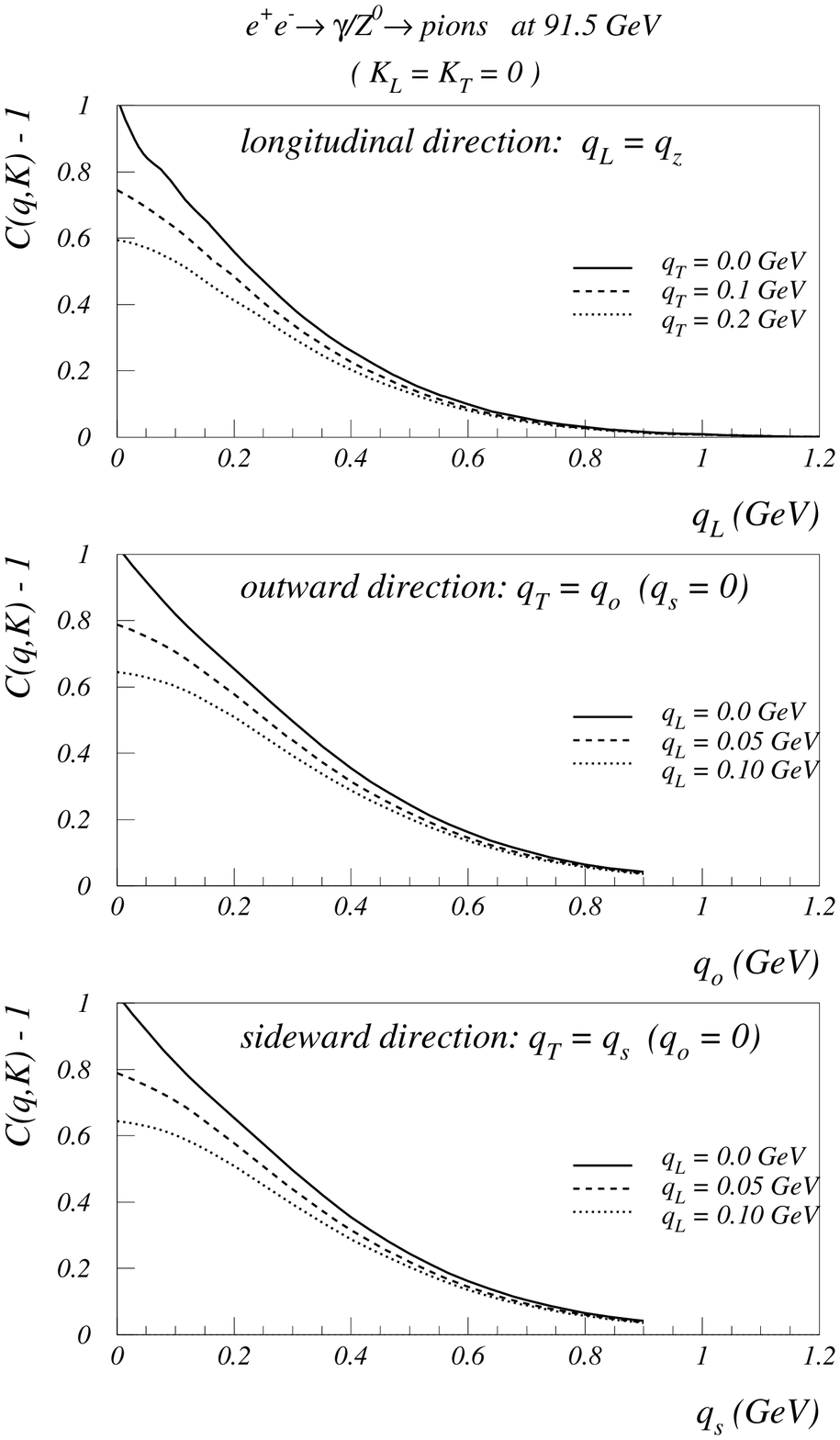} }
\end{minipage}
\epsfxsize=9.5cm
\begin{minipage}{80mm}
\centerline{ \epsfbox{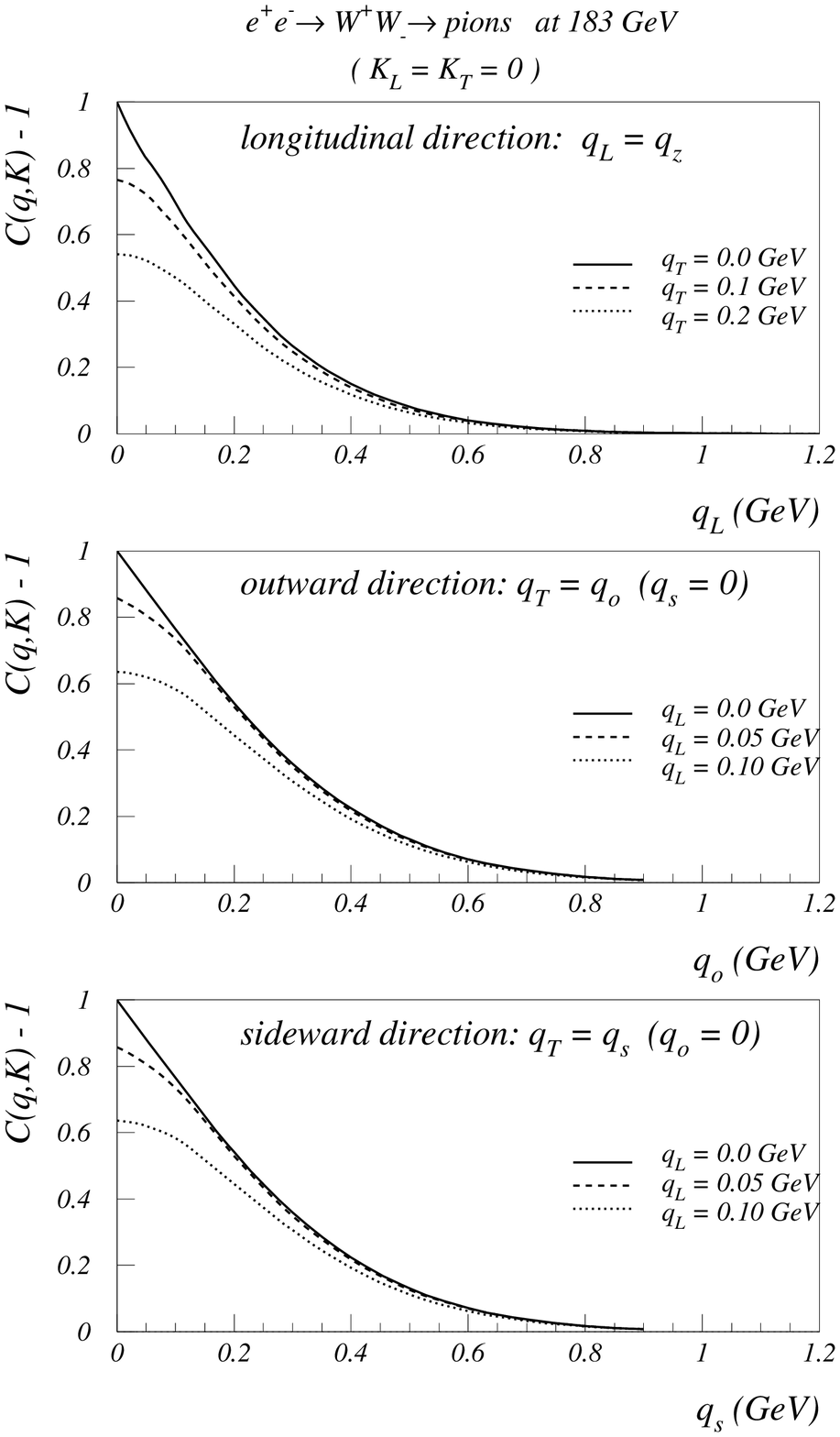} }
\end{minipage}
\caption{
The correlation function of same-sign pions for different values of 
the relative pair momentum $\bq$ for vanishing pair momentum $\bK$,
$C(q_z,q_s,q_o,\bbox{0})-1$.}
\label{fig5}
\end{figure}

\begin{description}
\item{a)}
In both cases, 
$e^+ e^-  \rightarrow   Z^0  \rightarrow q\bar{q} \rightarrow \pi's$
and
$e^+ e^-  \rightarrow   W^+W^-  \rightarrow q\bar{q}' \,q'\bar{q} \rightarrow 
\pi 's$,
we see {\it no} significant differences between the three relative
momentum directions $q_z, q_s, q_o$. Although, for a fixed direction
$q_i$, the intercept of the correlator at $q_i=0$ depends on the magnitude
of the momentum transverse to $q_i$, it looks the same for $q_i = q_z$,
$q_i = q_s$ or $q_i = q_o$.  
\item{b)}
The correlation function in the case 
$e^+ e^-  \rightarrow   W^+W^-  \rightarrow q\bar{q}' \,q'\bar{q} \rightarrow 
\pi 's$ is slightly narrower than that of
$e^+ e^-  \rightarrow   Z^0  \rightarrow q\bar{q} \rightarrow \pi's$.
Since the mean $q$ values correspond to the inverses of the typical
emission source sizes, this means that
the hadronic $W^+W^-$ decays reflect a larger
source size ($\simeq 1 $ fm) than the $Z^0$ decays ($\simeq 0.8$ fm), 
as we discuss later in the context of resonance effects.
\end{description}

The implication of observation a) is that the pion emission
appears essentially spherically symmetric with respect to the
three orthogonal directions $q_z, q_s, q_o$. This may appear to conflict
with the naive expectation that the source should appear much more 
elongated in the longitudinal $z$ direction than in the sideward 
and outward  $s,o$ directions, because of the large longitudinal 
momenta of the leading quark jets. However, as we pointed out 
before, in this model the pre-hadronic cluster formation is 
controlled by the `nearest-neighbor' criterion, so that only 
spatially adjacent partons with a mean separation $R_c \sim 0.8$ fm 
have a significant probability of coalescing and decaying into
pions and other hadrons. This local coalescence results naturally 
in a longitudinal-momentum ordering of particles as a function of 
their distance from the jet origin: particles further away tend to 
have higher momentum than those in the center. Since the Bose-Einstein 
effect is only apparent for identical particles with {\it similar} 
momenta, corresponding to small $q$, particles that are separated 
by many fermi at production are incapable of showing a significant 
enhancement because their momenta are so different. 

One may conclude from observation b) that parton `exogamy' in $W^+W^-$ 
decays~\cite{EG97} results in a space-time distribution of hadrons that 
is more spread out than in the case of $Z^0$ decays. This may again be 
understood as a consequence of increased efficiency of hadron formation 
in the $W^+W^-$ events, as we discussed before in the context of the 
pion multiplicity distributions in Fig.~\ref{fig4}. The identical pions 
emerging as products of parton-cluster decays have a longer distance 
correlation in  $W^+W^-$ events, because the partonic offspring of the 
overlapping $q\bar{q}$ dijets are enhanced mainly for low-momentum quanta 
in the central rapidity region, corresponding to significant `exogamous' 
coalescence of partons from different $W$'s with small relative 
momenta. As a consequence, the pion pair spectrum from $W^+W^-$
decays in Fig.~\ref{fig5}b is narrower than the one from $Z^0$ decays in 
Fig.~\ref{fig5}a, which translates into a larger effective emission radius 
for these pions.

\subsubsection{The pion correlator {\boldmath $C(q,K)$} for 
{\boldmath $K \ne 0$}}
\label{sec424}

The general features and physics interpretation of the $\bK$-dependence 
of the correlation function $C(\bq,\bK)$ have been studied in detail 
in \cite{H96a,H96,ursuli}. A manifest change in the shape of $C(\bq,\bK)$ 
as $\bK$ varies can have several origins, the two most important being (i) 
resonance decay contributions \cite{ursuli,schlei} and (ii) collective 
flow of the particle matter \cite{H96a}. Whereas pions from long-lived 
resonances are always present in high-energy collisions, collective motion
of the produced particles is a feature of heavy-ion reactions that produce 
high-density matter, but certainly is not an issue for the $e^+e^-$ 
collisions discussed here. In this subsection we do not distinguish 
the contributions from resonance decays, but show the $\bK$-dependence
of the correlation function including pions from long-lived resonances,
just as in the previous figures. We disentangle the effect of resonance 
decays in the next subsection.

Fig.~\ref{fig6} shows the correlation function $C(\bq,\bK) -1$ of same-sign 
pions for various values of the pair momentum $\bK = (K_L,\bK_\perp)$, 
where $K_L = K_z$ is the direction along the thrust axis and 
$K_\perp = |\bK_\perp|= \sqrt{K_x^2+K_y^2}$ the momentum transverse to it. 
The correlator is plotted as a function of one of the three Cartesian 
components of the relative momentum $\bq$, with the other two components
set to zero. The two main features are:

\begin{figure}[ht]
\vspace*{1cm}
\epsfxsize=9.5cm
\begin{minipage}{80mm}
\centerline{ \epsfbox{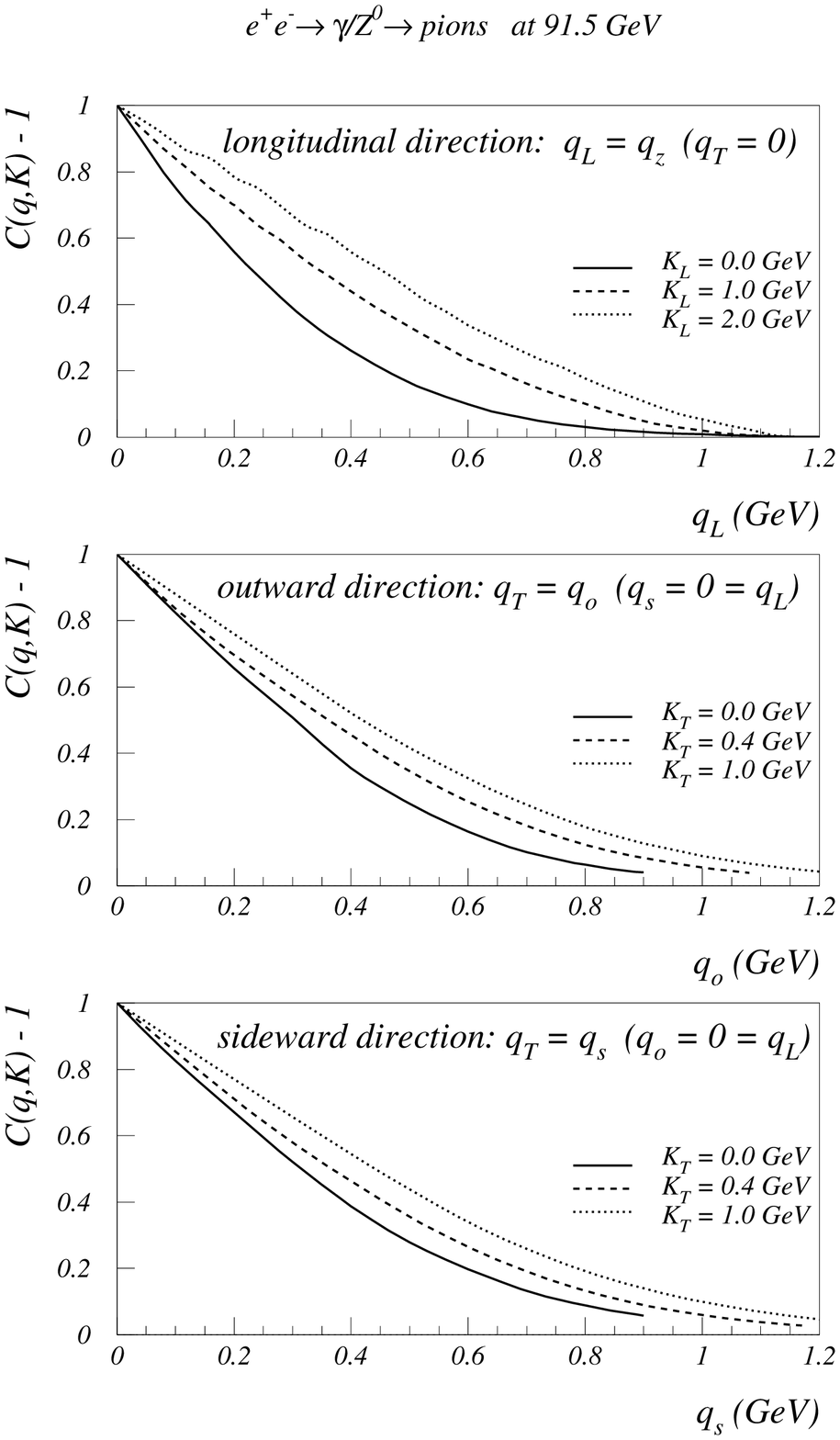} }
\end{minipage}
\epsfxsize=9.5cm
\begin{minipage}{80mm}
\centerline{ \epsfbox{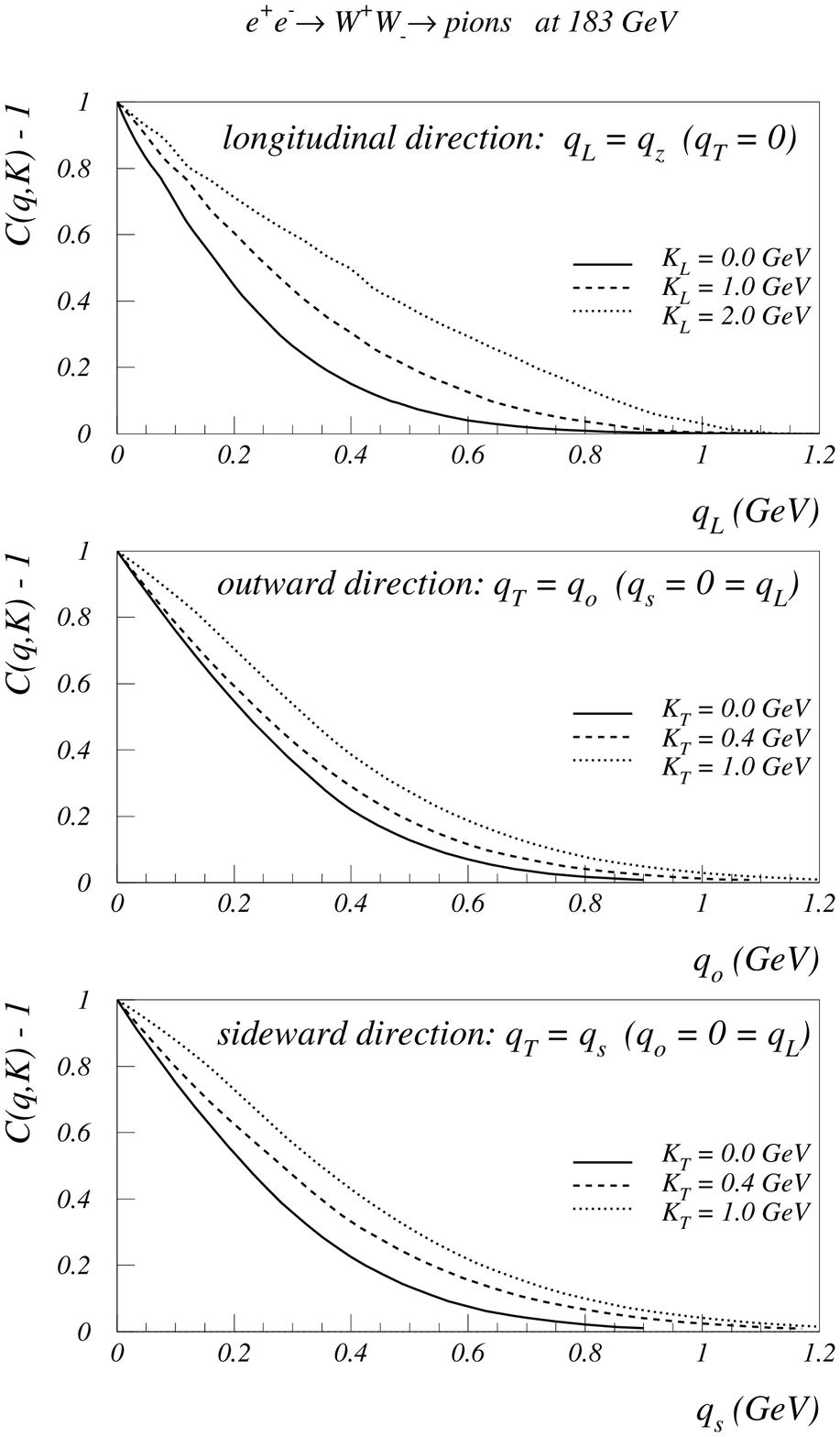} }
\end{minipage}
\caption{
The correlation function of same-sign pions $C(\bq,\bK)-1$ for various
values of the pair momentum $\bK = (K_L,\bK_\perp)$, where $K_L = K_z$ 
is the direction along the thrust axis and $K_\perp = |\bK_\perp|= 
\sqrt{K_x^2+K_y^2}$ is the momentum transverse to it. The correlators
are plotted against one component of the relative momentum, setting the 
two other components to zero.}
\label{fig6}
\end{figure}

\begin{description}
\item{a)}
The shape of the correlation function flattens and widens
as $K_L$ or $K_\perp$ are increased. However, the mean $q_{L,\perp}$
change by less than 10 \% when the $K_{L,T}$ values are varied 
from 0 to 1 GeV.
\item{b)}
The $\bK$-dependence is evidently spherically symmetric, i.e.,
the correlation function changes in the same way as $K_L$ or
$K_\perp$ is increased in the range from 0 to 1 GeV.
\end{description}

Point a) is a reflection of resonance-decay pions:
long-lived resonances with $\tau > 10$ fm
can travel for many fermis before decaying, which leads
to an exponential tail in the pion pair spectrum~\cite{ursuli}. 
This `life-time effect' is larger for small values of 
$\bK$ and damps out as $\bK$ is increased, since
the relative abundance of resonances is most pronounced
at small $\bK$.

Point b), on the other hand, is in accord with the fact that the
kinematics is approximately boost-invariant along the thrust axis
of the $e^+e^-$ collision, and the `local' character of the 
particle dynamics in our model. Neither
the parton shower evolution nor the parton-cluster
hadronization depend on the overall momentum $\bK$
relative to the $e^+e^-$ center-of-mass frame, as it is only
the kinematics, color and flavor of the near-by clustering partons
which at any given vertex determines locally the 
development of particle production.

\subsubsection{Effects of resonance decays on {\boldmath $C(q,K)$}}
\label{sec425}

Consider now a pair of identical pions with relative momentum $\bq$,
where one of the pions originates from a resonance of momentum $\bbox{p}$ 
with mass $m_r$ and decay width $\Gamma_r \sim \tau_r^{-1}$. Such a 
pair cannot contribute to the Bose-Einstein effect if $|\bq\cdot\bbox{p}| 
\gg m_r \Gamma_r$, which roughly implies that $|\bq| \gg \tau_r^{-1}$.
Since $\bq$ is inversely proportional to the spatial
dimension of the pion source, this means that resonances
represent a source of spatial extent of the order of $\tau_r^{-1}$.
Hence, such pions only can exhibit correlations if 
$|\bq| \le O(\tau_r^{-1})$, which for long-living resonances
($\tau_r \ge 10$ fm) requires $|\bq| \le 20$ MeV. This is less than 
the scale at which direct pions coming from the pre-hadronic clusters 
contribute, as seen in Table~\ref{T1}.
Therefore the pion correlation function $C(\bq,\bK)$ is narrowed
by the effects of the pion decay products of long-living resonances.

\begin{figure}[ht]\epsfxsize=14.5cm 
\centerline{\epsfbox{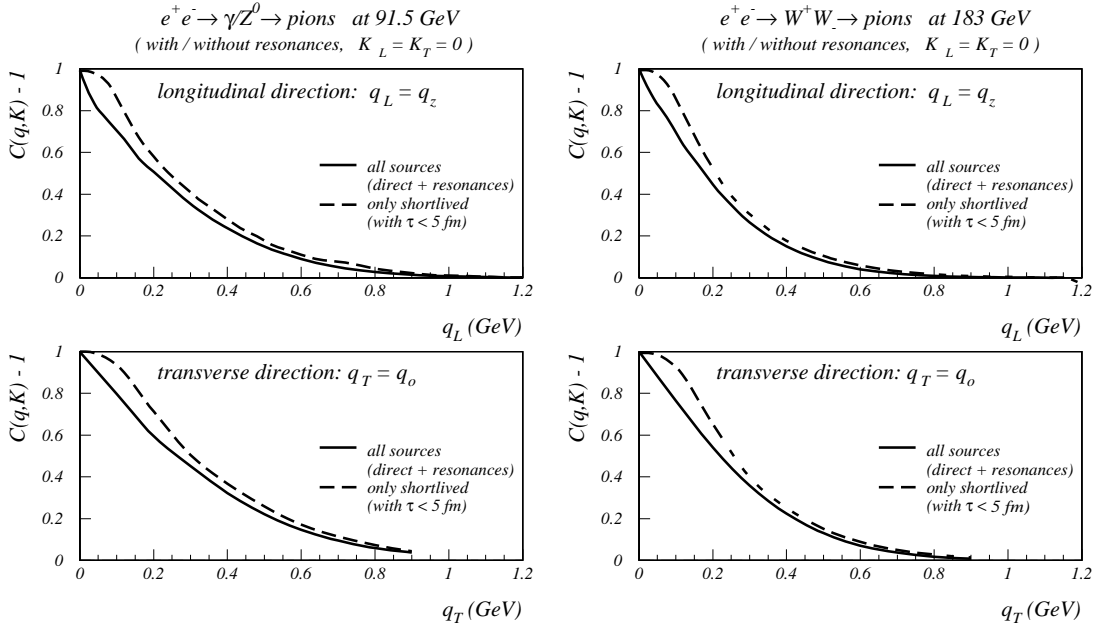}}
\caption{
The correlator $C(\bq,\bK)-1$ for $\bK=\bbox{0}$ with 
(solid curves) and without (dashed lines) the contributions from 
long-lived resonances.
}\label{fig7}
\end{figure}

To quantify the resonance narrowing and localize the pile-up of pion 
pairs where at least one comes from resonance decay, we have disentangled 
the pion emission sources of Table~\ref{T1} within our event simulation. 
In Fig.~\ref{fig7} we show again the correlator
$C(\bq,\bK)-1$ for $\bK = \bbox{0}$ in the two cases of 
$e^+ e^-  \rightarrow   W^+W^-  \rightarrow q\bar{q}' \,q'\bar{q} \rightarrow 
\pi$'s and $e^+ e^-  \rightarrow   Z^0  \rightarrow q\bar{q} \rightarrow 
\pi$'s. The full lines correspond to the correlator with all sources included,
as in the previous Fig.~\ref{fig6}, whereas the dashed curves
have the long-living resonance decay contributions removed.
One sees that the resonance decay pions make
a significant 20 - 30 \% contribution to the magnitude of $C - 1$ at
$q_L, q_T \, \lower3pt\hbox{$\buildrel <\over\sim$}\,50$ MeV,
corresponding to life-times $> 5$ fm.
%
\begin{figure}[ht]\epsfxsize=12.5cm 
\centerline{\epsfbox{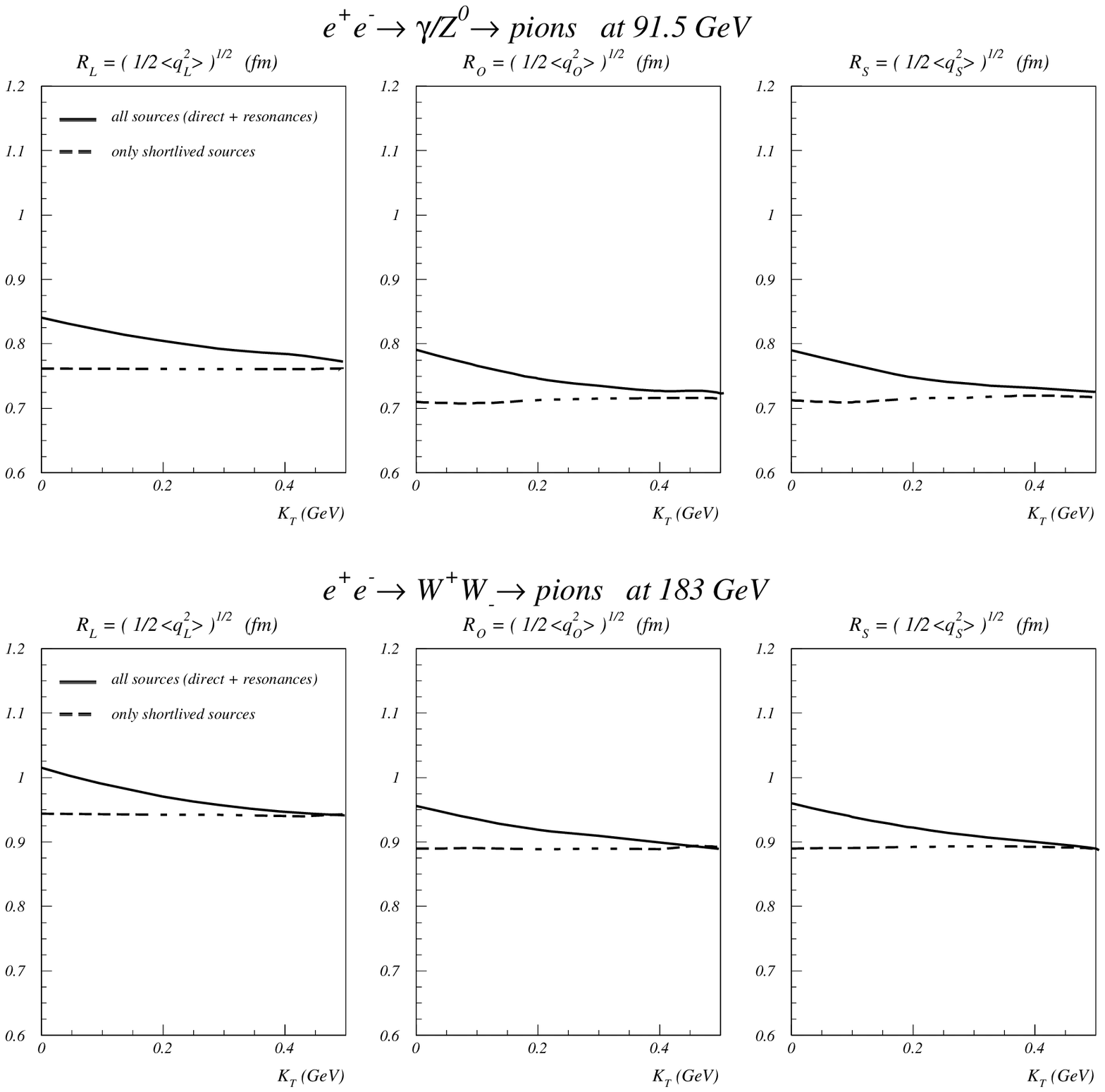}}
\caption{
The $\bK$ dependence of the effective pion source radii 
associated with the mean values of the components of 
$\bq = (q_L,q_s,q_o)$ in the longitudinal ($R_L$) and transverse
($R_o,R_s$) directions with respect to the thrust axis. The solid 
lines include all sources, whereas the dashed curves exclude
long-lived resonances.}
\label{fig8}
\end{figure}
%
Fig.~\ref{fig8} exhibits the $\bK$-dependence of the effective emission radii 
associated with the mean values of the components of $\bq = (q_L,q_s, q_o)$
in the longitudinal ($R_L$) and transverse ($R_o,R_s$) directions with 
respect to the thrust axis \cite{ursuli}:
\begin{equation}
R^2_i(\bK) 
\;=\; \frac{1}{2 \langle q_i^2 \rangle}
\;,\;\;\;\;\;\;\;\;\;\;\;\;\;\;
\langle q_i^2 \rangle \;=\;
\frac{ \int d q_i \, q_i^2 \,\left[ \,C(q_i,\bK) - 1 \right] }
     { \int d q_i \, \left[ \,C(q_i,\bK) - 1 \right] }
\;.
\end{equation}
Again, the full lines include all the sources in Table~\ref{T1}, whereas the 
dashed lines exclude the long-living resonances. It is evident that 
the effect of resonances is most pronounced for small values of $\bK$ 
and disappears with increasing $\bK$. This is expected as the abundance 
of pions from resonance decays is most prominent at small $\bK$.
Resonance decays thus induce a pair-momentum dependence of the HBT radii
\cite{ursuli,schlei}, with overall variations of the radii on the order of 
0.1 fm.

It is interesting to note that after switching off the resonance decay 
contributions the HBT radii shown in Fig.~\ref{fig8} exhibit no remaining 
pair-momentum dependence. Such a pair-momentum dependence would signal
the presence of position-momentum correlations in the source as, e.g., 
induced by collective flow or string breaking kinematics. No such 
correlations are visible here, not even along the thrust axis (see 
the left panels of Fig.~\ref{fig8} which show $R_L$ as a function of 
$K_\perp$). This is surprising because the inside-outside cascade
features of parton and hadron production in VNI should lead to appreciable
position-momentum correlations along the longitudinal axis defined by the
primary hard partons. We can think of two possible reasons for the fact
that they are not reflected in the longitudinal HBT radii shown here:
either they get largely averaged out by summing over many collision
events (which we think is unlikely), or they get covered up by the finite
size of the wave packet width $\sigma$ in the Bose-Einstein afterburner.
The fact that all HBT radii come out very close to $\sigma = 0.8$ fm lends
support to the second conjecture, although a final clarification of this 
issue has to await a comparison with calculations based on the ``classical'' 
version of the afterburner, as well as studies of the ``quantum''
version with different values of $\sigma$.

\section{Discussion}
\label{sec5}
%
We have discussed in this paper two possible algorithms for modelling
Bose-Einstein correlations in a Monte Carlo code for $e^+e^-$
annihilation into hadrons, that incorporates information from
perturbative QCD on the space-time evolution of parton showers and
a configuration-space criterion for hadronization. Afterburners
incorporating both the ``classical'' and ``quantum'' algorithms have been
applied to a model in which the hadron emission region is known
analytically. Standard tools for analyzing the sizes of
hadron emission regions have been applied to these model
calculations, and shown to reproduce successfully the parameters
of the model. The quantum algorithm has then been implemented as an
afterburner in the space-time parton-shower Monte Carlo, and
applied to $e^+e^- \rightarrow Z^0 \rightarrow hadrons$ and
$e^+e^- \rightarrow W^+W^- \rightarrow hadrons$. Exploratory
analyses have been presented of two-pion correlations in
longitudinal and transverse momenta, both with and without
resonance decays. The latter have been shown to modify significantly
the Gaussian behaviour that would otherwise have been expected, and
to cause a pair-momentum dependence of the extracted HBT radii, albeit
on a small scale of order 0.1 fm only. In the limit studied here
where the ``quantum'' afterburner was used with a fixed wave packet 
width $\sigma =0.8$ fm, resonance decays in fact induced the {\em only}
discernible $K_\perp$-dependence of the HBT radii.

The analysis of this paper has necessarily been incomplete, and we
conclude by listing some of the open questions that could be
addressed in any future work. It would be interesting to implement
the ``classical'' algorithm as an afterburner, and investigate the
similarities and differences with the quantum afterburner explored
in this paper. We are not in a position to express a definitive
theoretical
preference for one algorithm over the other. Within the context
of the quantum afterburner, we have assumed one particular value
of the Gaussian wave-packet size $\sigma$, and have not
explored the implications of varying this parameter. In this
connection, we should draw the reader's attention to the 
possibility that the
similarities between the correlations in the transverse and longitudinal
momenta may be related to the choice of $\sigma$, which we have not
attempted to optimize. This would involve an overall tuning of the
Monte Carlo to fit particle spectra, which we are currently not
in a position to complete\footnote{For this reason, it is not
possible currently to use this Monte Carlo to make a quantitative
estimate of the systematic uncertainties 
in the $W^{\pm}$ mass determination at LEP~2
due to colour reconnection or parton exogamy.}. Once
this is done, one could use the
Monte Carlo to address some of the physics issues that triggered
this investigation, including the possible effects of Bose-Einstein
correlations on measurements of the $W^{\pm}$ mass in hadronic
final states at LEP~2.

Despite the inevitable incompleteness of this work, we hope that the
ideas and investigations reported here may be useful in future studies
along the lines suggested above, either within the context of the
space-time approach to $e^+e^-$ annihilation into hadrons
used here, or within some other approach. The Bose-Einstein
afterburners studied here could also be implemented in Monte
Carlo codes for other interactions, including relativistic
heavy-ion collisions. We are currently studying how this work
could be advanced along these lines.

\bigskip

The authors acknowledge illuminating discussions with M. Gyulassy,
J. Sollfrank, S. Vance and W.A. Zajc. We thank C. Slotta for
assisting us with the figures. This work was supported in part by the 
Director, Office of Energy Research, Division of Nuclear Physics of the
Office of High Energy and Nuclear Physics of the U.S. Department of
Energy under Contract No. DE-FG02-93ER40764 and DE-AC02-76H00016.



\begin{thebibliography}{99}
\bibitem{strings}   
  B. Andersson, G. Gustafson, G. Ingelman, and T. Sj\"ostrand, 
  Phys. Rep. {\bf 97}, 33  (1983);
  B. Andersson, G. Gustafson and B. S\"oderberg, Nucl. Phys. {\bf B264}, 
  29 (1986).
\bibitem{clusters}   
  T. D. Gottschalk,  Nucl. Phys. {\bf B214}, 201 (1983)
  and Nucl. Phys. {\bf B227}, 413 (1983);
  B. R. Webber, Nucl. Phys. {\bf B238}, 492 (1984).
\bibitem{preconf}
  D. Amati and G. Veneziano, Phys. Lett. {\bf 83B}, 87 (1979);
  D. Amati, A. Bassetto, M. Ciafaloni, G. Marchesini, and G. Veneziano,
  Nucl. Phys. {\bf B173}, 429 (1980).
\bibitem{ms37} 
  J. Ellis and K. Geiger, Phys. Rev. D {\bf 52}, 1500 (1995);
  Nucl. Phys. {\bf A590}, 609 (1995).
\bibitem{GPZ}
  G. Gustafson, U. Pettersson, and P. Zerwas, Phys. Lett. B {\bf 209},
  90 (1988).
\bibitem{SK}
  T. Sj\"ostrand and V. A. Khoze, Phys. Rev. Lett. {\bf 72}, 28 (1994);
  Z. Phys. C {\bf 62}, 281 (1994).
\bibitem{GH}
  G. Gustafson and J. H\"akkinen, Z. Phys. C {\bf 64}, 659 (1994).
\bibitem{LL}
  L. L\"onnblad, Z. Phys. C {\bf 70}, 625 (1996).
\bibitem{EG96}
  J. Ellis and K. Geiger, Phys. Rev. D {\bf 54}, 1967 (1996). 
\bibitem{EG97} 
  J. Ellis and K. Geiger, Phys. Lett. B {\bf 404}, 230 (1997).
\bibitem{BEtheory}
  M. G. Bowler,  Z. Phys. C {\bf 29}, 617 (1985); 
  B. Andersson and W. Hoffmann, Phys. Lett. B {\bf 169}, 364 (1986);
  B. Andersson and M. Ringner, Phys. Lett. B {\bf 421}, 283 (1998);
  J. H\"akkinen  M. Ringner, Eur. Phys. J. {\bf C5}, 275 (1998).
\bibitem{LS95} 
  L. L\"onnblad and T. Sj\"ostrand, Phys. Lett. B {\bf 351}, 293 (1995).
\bibitem{LS97} 
  L. L\"onnblad and T. Sj\"ostrand, Eur. Phys. J. {\bf C2}, 165  (1998).
\bibitem{BEdata}
  TPC Collaboration, Phys. Rev. D {\bf 31}, 996 (1985);
  TASSO Collaboration, Z. Phys. C {\bf 30}, 355 (1986);
  MARK II Collaboration, Phys. Rev. D {\bf 39}, 39 (1989).
\bibitem{LEPBE} For analyses at the $Z^0$ peak, see:\\
  ALEPH Collaboration, Z. Phys. C {\bf 54}, 75 (1992);
  DELPHI Collaboration, Phys. Lett. B {\bf 268}, 201 (1992);
  OPAL Collaboration, Eur. Phys. J. {\bf C5}, 239 (1998) and references
        therein.
\bibitem{LEP2BE} For recent analyses in $W^+ W^-$ final states, see:\\
ALEPH Collaboration, paper 894;
DELPHI Collaboration, paper 288;
L3 Collaboration, paper 506;
OPAL Collaboration, paper 391
contributed to the {\sl International Conference on High-Energy Physics},
Vancouver, 1998.
\bibitem{Kunszt} Z. Kunszt {\sl et al.},
  {\sl Proceedings of the Workshop on Physics at LEP2},
  CERN Yellow Report 96-01, eds. G. Altarelli, T. Sj\"ostrand and F.
Zwirner, Vol. 1, 141 (1996).
\bibitem{vni} 
  K. Geiger,  Comp. Phys. Com. {\bf 104}, 70 (1997).
  The latest version of the computer program VNI can be obtained
              from {\it http://rhic.phys.columbia.edu/rhic/vni}.
\bibitem{transport}
K. Geiger, Phys. Rev. D {\bf 54}, 949 (1996).
\bibitem{CEO}
  B.A. Campbell, J. Ellis and K.A. Olive, Phys. Lett. B {\bf 235}, 325 (1990);
  and Nucl. Phys. {\bf B345}, 57 (1990).
\bibitem{EGK}
J. Ellis, K. Geiger and H. Kowalski, Phys. Rev. D {\bf 54}, 5443 (1996). 
\bibitem{GHI}
K. Geiger and R. Longacre, Heavy Ion Phys. {\bf 8}, 41 (1998).
\bibitem{GM}
K. Geiger and B. M\"uller, Heavy Ion Phys. {\bf 7}, 207 (1998);
K. Geiger, Phys. Rev. D {\bf 57}, 1895 (1998).
\bibitem{GS}
K. Geiger and D.K. Srivastava, Phys. Lett. B {\bf 422}, 39 (1998);
nucl-th/9806050; nucl-th/9808042.
\bibitem{newBE}
K. Fialkowski and R. Wit, Eur. Phys. J. {\bf C2}, 691 (1998); 
hep-ph/9805476; hep-ph/9810492; 
K. Fialkowski, R. Wit and J. Wosiek, Phys. Rev. D {\bf 58}, 094013 (1998);
A. Bialas and K. Zalewski, hep-ph/9806435;
S.V. Chekanov, E.A. De Wolf and W. Kittel, hep-ph/9809530.
\bibitem{H96a}
  U. Heinz, Nucl. Phys. {\bf A610}, 264c (1996).
\bibitem{WTH98}
  U.A. Wiedemann, B. Tom\'a\v{s}ik, and U. Heinz, Nucl. Phys.
  {\bf A638}, 475c (1998).
\bibitem{H96}
  U. Heinz, in {\it Correlations and Clustering Phenomena in Subatomic
  Physics}, eds. M.N. Harakeh, O. Scholten, and J.H. Koch, NATO 
  ASI Series B {\bf 359}, 137 (1997) (Plenum, New York); 
  nucl-th/9710065; hep-ph/9806512.
\bibitem{ZWSH97}
  Q.H. Zhang, U.A. Wiedemann, C. Slotta and U. Heinz,
  Phys. Lett. B {\bf 407}, 33 (1997).
\bibitem{Weal97} 
  U.A. Wiedemann, P. Foka, H. Kalechofsky, M. Martin,
  C. Slotta and Q.H. Zhang, Phys. Rev. C {\bf 56}, R614 (1997).
\bibitem{MKFW96}
  M. Martin, H. Kalechofsky, P. Foka and U.A. Wiedemann, 
  Eur. Phys. J. {\bf C2} (1998) 359.
\bibitem{WEHK98}  U.A. Wiedemann, J. Ellis, U. Heinz, and K. Geiger, 
  to appear in {\it CRIS'98: Measuring the size of things in the 
  Universe: HBT interferometry and heavy ion physics},
  Catania, June 8-12, 1998, eds. S. Costa {\sl et al.}, (World
  Scientific, Singapore, 1998), nucl-th/9808043.
\bibitem{Zajc}
  W.A. Zajc, in {\it Particle Production in Highly Excited Matter}, 
  eds. H.H. Gutbrod and J. Rafelski, NATO ASI Series B {\bf 303}, 435 (1993)
  (Plenum, New York).
\bibitem{AR97}
  B. Andersson and M. Ringner, Nucl. Phys. {\bf B513}, 627 (1998);
  Phys. Lett. B {\bf 421}, 283 (1998).
\bibitem{AHR97}
  D.V. Anchishkin, U. Heinz, and P. Renk, Phys. Rev. C {\bf 57}, 1428 (1998).
\bibitem{W98}
  U.A. Wiedemann, Phys. Rev. C {\bf 57}, 3324 (1998).
\bibitem{S73} 
  E. Shuryak, Phys. Lett. {\bf 44B}, 387 (1973);
  Sov. J. Nucl. Phys. {\bf 18}, 667 (1974).
\bibitem{P84}
  S. Pratt, Phys. Rev. Lett. {\bf 53}, 1219 (1984); Phys. Rev. D {\bf 33},
  72 (1986).
\bibitem{CH94} 
  S. Chapman and U. Heinz, Phys. Lett. B {\bf 340}, 250 (1994).
\bibitem{GKW79}
  M. Gyulassy, S.K. Kauffmann, and L.W. Wilson, Phys. Rev. C {\bf 20}, 
  2267 (1979).
\bibitem{MV97} D. Mi\'skowiec and S. Voloshin, nucl-ex/9704006.
\bibitem{ZSH98}
  Q.H. Zhang, P. Scotto, and U. Heinz, nucl-th/9805046, to be published in
        Phys. Rev. C {\bf 58} (1st Dec., 1998).
\bibitem{B94}
  G. Bertsch, Phys. Rev. Lett. {\bf 72}, 2349 (1994).
\bibitem{prattrev}
  S. Pratt, in {\it Quark-Gluon Plasma 2}, ed. R.C. Hwa
  (World Scientific, Singapore, 1995), p. 700.
\bibitem{PGG90}
  S. Padula, M. Gyulassy, and S. Gavin, Nucl. Phys. {\bf B329}, 357 (1990).
\bibitem{CZ97}
  T. Cs\"org\H{o} and J. Zim\'anyi, Phys. Rev. Lett. {\bf 80}, 916 (1998);
  J. Zim\'anyi and T. Cs\"org\H{o}, hep-ph/9705432.
\bibitem{LEP1}
  T. Hebbeker,  Phys. Rep. {\bf 217}, 69 (1992);
  S. Bethke and J. E. Pilcher, Ann. Rev. Nucl. Part. Sci. {\bf 42}, 251 (1992).
\bibitem{LEP2}
  A. Ballestrero {\it et al.}, J. Phys. G {\bf 24}, 365 (1998).
\bibitem{haywood}
  S. Haywood, Rutherford-Appleton preprint RAL-94-074, (1994).
\bibitem{evgens} 
  For example:
  G. Marchesini, B. R. Webber, G. Abbiendi, I. G. Knowles, M. H. Seymor 
  anp L. Stanco, Comp. Phys. Com.  {\bf 67}, 465 (1992);
  L. L\"onnblad, Comp. Phys. Com. {\bf 71}, 15 (1992);
  T. Sj\"ostrand, Comp. Phys. Com. {\bf 82}, 74 (1994).
\bibitem{ms3942} 
  K. Geiger, Phys. Rev. D {\bf 56}, 2665 (1997).
\bibitem{marchesi}
  G. Marchesini, L. Trentadue and G. Veneziano, Nucl. Phys. {\bf B181}, 335
  (1981); K. Konishi, CERN-TH.2853 (1980).
\bibitem{BEresonances}
  P. Grassberger, Nucl. Phys. {\bf B120}, 231 (1977);
  M. Bowler, Z. Phys. C {\bf 46}, 305 (1990).
\bibitem{ursuli}
  U. A. Wiedemann and U. Heinz, Phys. Rev. C {\bf 56}, 610 and 3265 (1997).
\bibitem{schlei}
  B.R. Schlei, U. Ornik, M. Pl\"umer and R. M. Weiner, Phys. Lett. B 
  {\bf 293}, 275 (1992); 
  J. Bolz, U. Ornik, M. Pl\"umer, B.R. Schlei and R. M. Weiner, 
  Phys. Lett. B {\bf 300}, 404 (1993); Phys. Rev. D {\bf 47}, 3860 
  (1993).
\end{thebibliography}
\end{document}